\documentclass[sigconf]{acmart}

\usepackage{amsmath,epsfig} 
\usepackage{soul}
\usepackage{makecell}
\usepackage{array}  
\usepackage{algorithmic}
\usepackage{algorithm,bbding,wasysym }
\usepackage{subfigure}
\usepackage{multirow}

\usepackage{amsmath,amssymb}
\usepackage{booktabs}
\usepackage{bbding}
\usepackage{pifont}

\AtBeginDocument{%
  \providecommand\BibTeX{{%
    \normalfont B\kern-0.5em{\scshape i\kern-0.25em b}\kern-0.8em\TeX}}}

\copyrightyear{2020} 
\acmYear{2020} 
\setcopyright{acmcopyright}\acmConference[CIKM '20]{Proceedings of the 29th ACM International Conference on Information and Knowledge Management}{October 19--23, 2020}{Virtual Event, Ireland}
\acmBooktitle{Proceedings of the 29th ACM International Conference on Information and Knowledge Management (CIKM '20), October 19--23, 2020, Virtual Event, Ireland}
\acmPrice{15.00}
\acmDOI{10.1145/3340531.3411869}
\acmISBN{978-1-4503-6859-9/20/10}

\begin{document}

%%
%% The "title" command has an optional parameter,
%% allowing the author to define a "short title" to be used in page headers.
\title{Sequential Recommender via Time-aware Attentive Memory Network}

%%
%% The "author" command and its associated commands are used to define
%% the authors and their affiliations.
%% Of note is the shared affiliation of the first two authors, and the
%% "authornote" and "authornotemark" commands
%% used to denote shared contribution to the research.
%\author{Ben Trovato}
%\authornote{Both authors contributed equally to this research.}
%\email{trovato@corporation.com}
%\orcid{1234-5678-9012}
%\author{G.K.M. Tobin}
%\authornotemark[1]
%\email{webmaster@marysville-ohio.com}
%\affiliation{%
%  \institution{Institute for Clarity in Documentation}
 % \streetaddress{P.O. Box 1212}
%  \city{Dublin}
 % \state{Ohio}
%  \postcode{43017-6221}
%}

\author{Wendi Ji}
\affiliation{\institution{East China Normal University}}
\email{wendyg8886@gmail.com}

\author{Keqiang Wang}
\authornote{Corresponding author.}
\affiliation{\institution{Pingan Health Technology}}
\email{wangkeqiang265@pingan.com.cn}

\author{Xiaoling Wang}
\affiliation{
 \institution{East China Normal University}
 \institution{Tongji University}}
\email{xlwang@cs.ecnu.edu.cn}

\author{Tingwei Chen}
\authornotemark[1]
\affiliation{
 \institution{Liaoning University}}
\email{tingwei.chen@durham.ac.uk}

\author{Alexandra Cristea}
\affiliation{
\institution{Durham University}}
\email{alexandra.i.cristea@durham.ac.uk}

%\author{Keqiang Wang}
%\affiliation{
% \institution{Pingan Health Technology}}
%\email{wangkeqiang265@pingan.com.cn}

%\author{Xiaoling Wang}
%\affiliation{
 %\institution{East China Normal University, Tongji University}}
%\email{xlwang@cs.ecnu.edu.cn}

%\author{Tingwei	Chen}
%\affiliation{
% \institution{Liaoning University}}
%\email{tingwei.chen@durham.ac.uk}

%\author{Alexandra Cristea}
%\affiliation{
% \institution{Durham University}}
%\email{alexandra.i.cristea@durham.ac.uk}
% \streetaddress{Rono-Hills}
% \city{Doimukh}
% \state{Arunachal Pradesh}
 %\country{India}}

%\author{Charles Palmer}
%\affiliation{%
%  \institution{Palmer Research Laboratories}
%  \streetaddress{8600 Datapoint Drive}
%  \city{San Antonio}
%  \state{Texas}
%  \postcode{78229}}
%\email{cpalmer@prl.com}

%\author{John Smith}
%\affiliation{\institution{The Th{\o}rv{\"a}ld Group}}
%\email{jsmith@affiliation.org}

%\author{Julius P. Kumquat}
%\affiliation{\institution{The Kumquat Consortium}}
%\email{jpkumquat@consortium.net}

%%
%% By default, the full list of authors will be used in the page
%% headers. Often, this list is too long, and will overlap
%% other information printed in the page headers. This command allows
%% the author to define a more concise list
%% of authors' names for this purpose.
%\renewcommand{\shortauthors}{Trovato and Tobin, et al.}

%%
%% The abstract is a short summary of the work to be presented in the
%% article.

\begin{abstract}
%In the era of information overload, r
Recommendation systems aim to assist users to discover most preferred contents from an ever-growing corpus of items.
Although recommenders have been greatly improved by deep learning, they still face several challenges: % in sequential recommendation: 
%, which is a recent research hotspot. 
%A common recipe is to use recurrent and attentive neural networks to capture the sequential dynamic.
%However, user behaviors are much more complex than words in sentences, where the correlations between items are not only determined by their positions but also  {\bf time}. 
(1) Behaviors are much more complex than words in sentences, so traditional attentive and recurrent models have limitations capturing the temporal dynamics of user preferences.
(2) The preferences of users are multiple and evolving, so it is difficult to integrate long-term memory and short-term intent.
%of between items in an interaction sequence are not only determined by their positions but also  {\bf time}.
%

In this paper, we propose a temporal gating methodology to improve attention mechanism and recurrent units, so that temporal information can be considered in both information filtering and state transition.
Additionally, we propose a hybrid sequential recommender, named {\bf M}ulti-hop {\bf T}ime-aware {\bf A}ttentive {\bf M}emory network ({\bf MTAM}), to integrate long-term and short-term preferences.
We use the proposed time-aware GRU network to learn the short-term intent and maintain prior records in user memory.
We treat the short-term intent as a query and design a multi-hop memory reading operation via the proposed time-aware attention to generate user representation based on the current intent and long-term memory.
Our approach is scalable for candidate retrieval tasks and can be viewed as a non-linear generalization of latent factorization for dot-product based Top-K recommendation.
%Our approach can be viewed as a non-linear generalization of latent factorization for dot-product based Top-K recommendation, which makes it scalable for candidate retrieval tasks. % under stringent latency requirements.
Finally, we conduct extensive experiments on six benchmark datasets and the experimental results demonstrate the effectiveness of our MTAM and temporal gating methodology.

\end{abstract}

%%
%% The code below is generated by the tool at http://dl.acm.org/ccs.cfm.
%% Please copy and paste the code instead of the example below.
%%
%\begin{CCSXML}
%<ccs2012>
 %<concept>
  %<concept_id>10010520.10010553.10010562</concept_id>
  %<concept_desc>Computer systems organization~Embedded systems</concept_desc>
 % <concept_significance>500</concept_significance>
 %</concept>
 %<concept>
  %<concept_id>10010520.10010575.10010755</concept_id>
 % <concept_desc>Computer systems organization~Redundancy</concept_desc>
 % <concept_significance>300</concept_significance>
% </concept>
 %<concept>
 % <concept_id>10010520.10010553.10010554</concept_id>
 % <concept_desc>Computer systems organization~Robotics</concept_desc>
%  <concept_significance>100</concept_significance>
 %</concept>
% <concept>
 % <concept_id>10003033.10003083.10003095</concept_id>
 % <concept_desc>Networks~Network reliability</concept_desc>
 % <concept_significance>100</concept_significance>
 %</concept>
%</ccs2012>
%\end{CCSXML}
%\ccsdesc[300]{Computer systems organization~Redundancy}
%\ccsdesc{Computer systems organization~Robotics}
%\ccsdesc[100]{Networks~Network reliability}

%%
%% Keywords. The author(s) should pick words that accurately describe
%% the work being presented. Separate the keywords with commas.

%% A "teaser" image appears between the author and affiliation
%% information and the body of the document, and typically spans the
%% page.

%%
%% This command processes the author and affiliation and title
%% information and builds the first part of the formatted document.
\maketitle
\vspace{-5pt}
\section{Introduction}
In large-scale recommendation systems, it is challenging to retrieval a set of most relevant items for a user given her/his interaction history from tens or hundreds of millions of items.
%And the responses have to be returned within hundreds of milliseconds, which make recommender systems need to balance between accuracy and efficiency.
A common recipe to handle the huge amount and sparsity of item corpus is matrix factorization, which facilitates efficient approximate k-nearest neighbor searches via resorting to the inner product of user representation and item representation 
%koren2008factorization,
\cite{su2009survey,rendle2008online,baltrunas2011matrix,rendle2009bpr}.
%In recent years, deep learning has achieved great progress in recommender systems.
%Neural recommenders leverage DNNs \cite{he2017neural}, RNNs \cite{Hidasi2015SessionbasedRW}, attentive networks \cite{liu2018stamp} and memory networks \cite{ebesu2018collaborative} to break the limitation of traditional latent factor models of modeling the nonlinear correlations between users and items. 

\begin{figure*} 
 \setlength{\belowcaptionskip}{-15pt} 
  \setlength{\abovecaptionskip}{0pt}   
  \centering
  \includegraphics[width=1.9\columnwidth,height=0.3\textwidth]{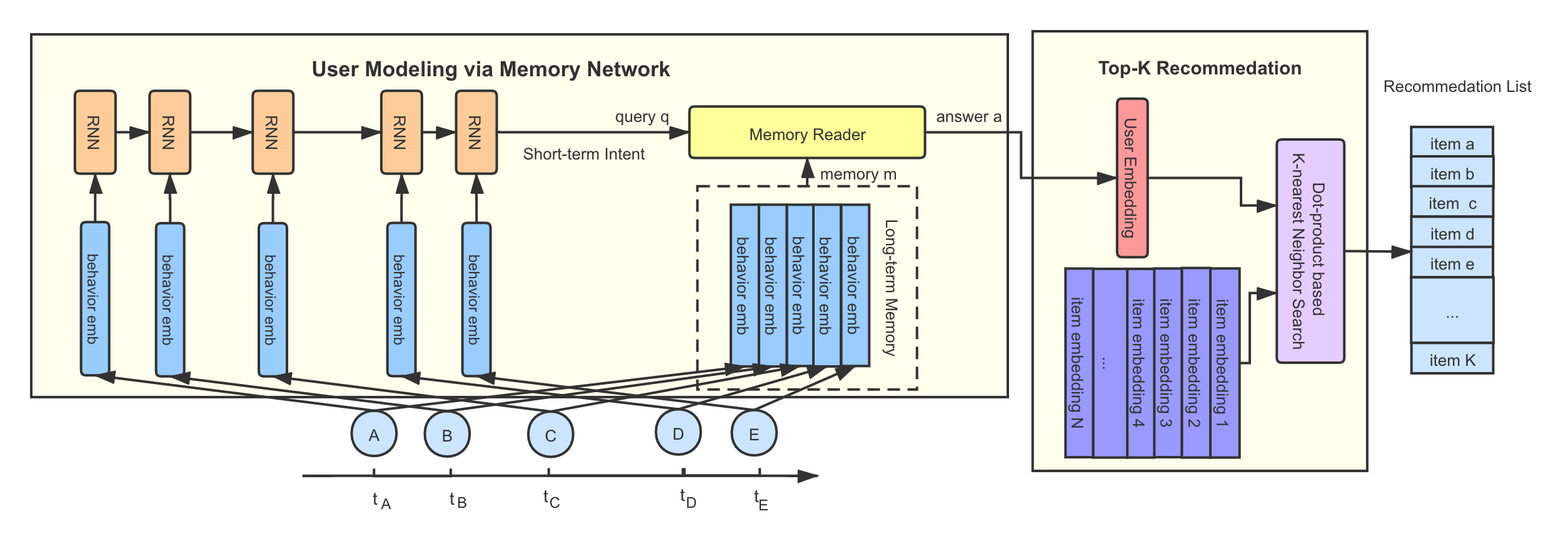}
  \caption{The general framework of MTAM for Top-K recommendation. The left part shows the user modeling module, which encodes user's short-term intent and long-term memory into the final user embedding via a memory network. And the right part shows the Top-K recommendation module, which produces a ranking list over all items by dot-product based K-nearest neighbor search.}
  \label{fig:general_framework}  
\end{figure*}

Typically, there are two stages in an industrial recommendation system: candidate generation and ranking \cite{covington2016deep,beutel2018latent,yi2019sampling}.
At the candidate generation stage, time-efficient neural nominators retrieval hundreds of candidates from a large corpus of items.
The candidates then re-ranked by a fully-blown neural ranking model.
The main difference between the two stages is that a ranking model can serve as a discriminator which predicts $score(u,i)$ (the preference of user $u$ on item $i$) on a small candidate set, while candidate generators are required to generate the representation of a target user which can be used in k-nearest neighbor searches.
%Because in corpus with millions of items, it is not computationally capable to exam the $score(u,i)$ of all items with a complex neural model. 
%Meanwhile, the predicted user representation can be directly used to compute the user's preferences for all items via inner product in large-scale retrieval systems.
As shown in Figure \ref{fig:general_framework}, we focus on the candidate generation stage which determines the ceiling performance of recommendation and treat it as a user modeling task .

%To predict , recommenders should take comprehensively account of the various, evolving and personalized preferences of users.
Analogous with words of sentences, a user's interactions with items naturally form a behavior sequence.
With the quick development of deep learning, many recent researches have built recurrent and attentive models to capture the sequential property of user behaviors \cite{beutel2018latent,Hidasi2015SessionbasedRW,li2017neural}.
% when predicting what is the next item a target user will buy.

A vital challenge of user modeling is that user behaviors are much more complex than words.
Some context features of a behavior, like category, action and text information, can be incorporated by injecting feature embeddings into the item embedding.
However, the temporal feature is related to a pair of behaviors.
Two behaviors within a short interval intuitively tend to be more relevant than two behaviors within a long interval.
Therefore, classical structures of recurrent and attentive networks need to be upgraded to model the temporal dynamics of sequential data better.
Some recent researches improve GRU or LSTM units by adding time gates to capture the temporal information of user behavior sequences \cite{zhu2017next,chen2019dynamic,yu2019adaptive}.
However, the temporal information is often ignored in attention mechanism.
Originated from the NLP field, attention mechanism takes a weighted sum of all components and focuses only on the information related to a query. 
Many recent neural recommenders utilize attention mechanism to filter diverse user interests by concentrating on relevant behaviors and eliminating irrelevant ones to predict a user's future action \cite{yu2019multi,tran2019signed,liu2018stamp,li2017neural,zhou2019deep,Feng2019Deep}. 
Among these attentive neural recommenders, researches \cite{li2017neural,Feng2019Deep} assign weights to compressed hidden states to build attentive RNN models; researches \cite{liu2018stamp,tran2019signed} assign weights to historical records to build attentive memory networks; and researches \cite{yu2019multi,zhou2019deep} assign weights to hidden variables to build deep attentive feed-forward networks.
But when calculating the correlation between two behaviors, the time interval between them has not been taken into consideration by most of the previous approaches.
One recent research \cite{li2020time} copes with this challenge by directly adding the clipped relative time intervals to the dot-product of item embeddings when calculating the attention weights.
However, how to upgrade attention mechanism to model the temporal patterns of sequences in a more fine-grained way is still a problem. 

To overcome the aforementioned limitations of neural networks, we propose a novel temporal gating methodology to upgrade attention mechanism and recurrent units by taking advantage of the time-aware distance between interactions in the task of user modeling.
Inspired by the gated mechanisms in LSTM and GRU, we introduce a temporal controller to encode the temporal distance between two interactions into a gate.
We then propose a novel time-aware attention by equipping scaled dot-product attention with the proposed temporal gate.
When calculating the relationship between two interactions, the time-aware attention kernel is capable to take the time interval into account.
Meanwhile, for better short-term preference modeling, we utilize the proposed temporal gate in GRU to control how much past information can be transferred to future states, named T-GRU.
Different from previous time-aware recurrent recommenders which aim to capture both the short-term and long-term interests \cite{zhu2017next,chen2019dynamic,yu2019adaptive}, we only aim at short-term intent by filtering out irrelevant past information with the temporal gate.
%hybrid recommender 

%The preferences and interests of users are both multiple and dynamic.
Additionally, both long-term preferences and short-term intents determine the behaviors of users.
In this paper, we view user behaviors as a decision making program in Memory Network and propose a {\bf M}ulti-hop {\bf T}ime-aware {\bf A}ttentive {\bf M}emory (MTAM for short) network based on the proposed T-GRU and time-aware attention, which is illustrated in Figure \ref{fig:general_framework}.
MTAM first utilizes T-GRU to capture the short-term intent of a user and maintains a fixed length history behaviors in a memory matrix (memory m) to store her/his long-term preference.
Inspired by the memory retrieval procedure in human mind, MTAM treats the short-term intent as a query $q$ to search throughout the long-term memory $m$.
The searching procedure of MTAM is to find a continuous representation for $m$ and $q$ via time-aware attention.
When making recommendations, MTAM processes the searching procedure for multi-hops to output an answer $a$, which is a comprehensive representation of the target user.
The time-aware attention mechanism of MTAM provides an effective manner to learn the temporal dynamics of behavior sequences by crediting the different contributions of prior interactions to the current decision.
%generate a multiple, evolving and personalized embedding as user representation
% in candidate generation stage.
The proposed MTAM can be viewed as a non-linear generalization of factorization techniques and applied in large-scale retrieval systems.
We will show experimentally that the temporal gating methodology and the multi-hop structure are crucial to good retrieval performance of MTAM in the Top-K recommendation task on 6 real-world datasets.

%MTAM can be viewed as a non-linear generalization of factorization techniques.
%the engages an explicit memory to store long-term preferences and multi-hop time-aware attention mechanism to read the memory for a short-tern intent, 
%  takes account of the various, evolving and personalized preferences of users.

%propose a novel time-aware attention mechanism by adding a temporal controller in scaled dot-product attention. 
In summary, the main contributions of the paper can be illustrated as follows:
 \begin{itemize}
\item We improve attention mechanism and recurrent unit via a novel temporal gating methodology to capture the temporal dynamics of users' sequential behaviors.
%better in user modeling tasks. %They are time-aware attention and T-GRU.
\item We propose a novel multi-hop time-aware memory neural network for sequential recommendation. MTAM treats the output of T-GRU as short-term intent and reads out the long-term memory effectively via time-aware attention.
%To the best of our knowledge, this is the first attempt to build neural memory recommender for candidate generation stage.
To the best of our knowledge, MTAM is the first memory network which takes the time-aware distances between items into account.

%Both proposed model are trained as learning-to-rank problem to cater to Top K recommendation.
\item We compare our model with state-of-the-art methods on six real-world datasets.
The results demonstrate the performance of Top-K recommendation is obvious improved via adding the temporal gate into recurrent unit and attention mechanism.
And compared with encoding the user representation with the weighted sum of recurrent hidden states, MTAM is able to leverage user historical records in a more effective manner.
\end{itemize}

\vspace{-12pt}
\section{Related Work}
Our work focuses on the candidate generation stage and is essentially a memory-augmented sequential recommender. We will review the related works in three directions.
\vspace{-5pt}
\subsection{Candidate Generation and Ranking}
In general, an industrial recommendation system consists of two stages: candidate generation and candidate ranking \cite{covington2016deep,beutel2018latent,yi2019sampling}.
The candidate generation (a.k.a. retrieval or nomination) stage aims to provide a small set of related items from a large corpus under stringent latency requirements \cite{covington2016deep,yi2019sampling}.
Then, the candidate ranking model reranks the retrieved items based on click-through rate (CTR for short), rating or score \cite{Feng2019Deep,wu2019dual,tran2019signed}.
In the retrieval stage, recommenders have to face the computational barriers of full corpus retrieval. 
A common recipe for candidate retrieval is modeling the user-item preference as the dot-product of the low dimensional user representation and item representation, such as matrix factorization\cite{koren2009matrix,rendle2009bpr} and neural recommenders \cite{covington2016deep,beutel2018latent,li2017neural}.
%, MLP recommender \cite{covington2016deep}, recurrent neural recommender \cite{beutel2018latent} and attentive neural recommender \cite{li2017neural}.
However, the dot-product correlation limits the capability of neural recommenders. 
%There are many more expressive CTR or score prediction models in ranking stage via modeling the correlations between user past behaviors and candidate items,
To learn deeper non-linear relationships between a target user and candidate items,
some more expressive models have been proposed for the ranking stage, 
such as neural collaborative filtering \cite{he2017neural}, deep interest network \cite{zhou2019deep}, SLi-Rec \cite{yu2019adaptive} and user memory network \cite{chen2018sequential}.
In this paper, we focus on the candidate generation stage and the purposed MTAM be viewed as a non-linear generalization of factorization techniques.
\vspace{-5pt}
\subsection{Sequential Recommenders}
Analogous with words of sentences in natural language processing (NLP), a user's interactions with items naturally form a behavior sequence.
In recent years, deep neural networks have achieved continuous improvements in NLP \cite{gehring2017convolutional, Mikolov2013EfficientEO, Devlin2018BERTPO, vaswani2017attention}, which prompt a series of explorations in applying neural networks in sequential recommendation.
Research\cite{Hidasi2015SessionbasedRW}, the first stab at employing RNN-based models in session-based recommendation, uses %Gated Recurrent Unit 
GRU to model the click sequences and improves the CTR prediction by taking the sequential characteristics into consideration.
Additionally, a user's purchase decision is both determined by her/his long-term stable interests and short-term intents \cite{devooght2017long,an2019neural,yu2019multi}.
Researches\cite{hu2017diversifying,li2017neural,liu2018stamp,zhou2019deep} combine RNNs with attention mechanism to learn the preference evolution of users.
%However, RNN-based models is natural not suitable to learning the long-term dependencies of sequences, even though gated techniques and attention mechanism have be applied \cite{ke2018sparse}.

However, although traditional attentive and recurrent models have shown excellent performance to model the sequential patterns of user behaviors, they only consider the orders of objects without the notion of the temporal information .
The time intervals between interactions are important to capture the correlations between interactions.
Research \cite{zhu2017next} improves LSTM by proposing some temporal gates to capture both long-term and short term preferences of users.
Researches \cite{chen2019dynamic, yu2019adaptive} further propose two time-aware recurrent units.
The main difference between our proposed T-GRU and previous models is we only use the time interval between adjacent interaction to control how much past information can be transferred to future states, while previous researches apply temporal gates to control both previous information and current content.
Furthermore, how to capture the temporal context in attention mechanism is still not well explored. 
Recent research \cite{li2020time} explores the influence of different time intervals on next item prediction and proposes a time-aware self-attentive model.
It treats time intervals as special positions and solves it by adding clipped intervals to the dot-product of item embeddings.
In this work, we update attention mechanism by a gating technique which helps to capture the non-linear temporal differences between interactions.
\vspace{-5pt}
\subsection{Memory-augmented Recommenders}
%The interests or preferences of users can be both stable and evolving.
The essential of user modeling is learning the dependencies of behaviors.
RNN-based models encode user's previous behaviors into hidden states.
Although attention mechanism helps recurrent networks to learn long-term dependencies by concentrating on the relevant states \cite{wang2019collaborative,li2017neural}, it fails to distinguish the different role that each item plays in prediction.
To tackle with this challenge, external memory networks have been proposed in recent years to store and manipulate sequences effectively \cite{graves2014neural,graves2016hybrid}, which have been successfully adapted to NLP tasks, such as question answering \cite{kumar2016ask}, knowledge tracing \cite{zhang2017dynamic}, translation \cite{maruf2018document} and dialogue systems \cite{wu2019global}.
Several recent researches propose memory-augmented neural networks for recommendation to leverage users' historical behaviors in a more effective manner \cite{chen2018sequential, liu2018stamp, ebesu2018collaborative, tran2019signed}.
They introduce an external user memory to maintain users' historical information and use attention mechanism to design memory reading operations.
Among these researches, \cite{chen2018sequential, ebesu2018collaborative, tran2019signed} treat the target item as the query, while \cite{liu2018stamp} treats the last item in the interaction sequence as the query.

There are two main differences between these existing researches and the proposed MTAM.
(1) Taking advantage of the proposed time-aware attention, MTAM is the first memory network which takes the temporal context of interactions into consideration. 
(2) Apart from \cite{liu2018stamp,wang2019collaborative}, most previous memory-augmented recommenders focus on the candidate ranking stage, which aim to predict $score(u,i)$ in ranking stage. Meanwhile, MTAM focuses on the candidate retrieval stage, which aims to generate the representation of a target user merely based on the her/his historical records.
%According to this temporal characteristic of the relationship between users and items, relevant researches of recommendation systems can be classified into two main paradigms: General Recommendation and Sequential Recommendation.

%aims to explore the preferences, habits and present intents from the users' historical behaviors.
\vspace{-5pt}
\section{Overall Framework}
We first give the formal notations that will be used in this paper and define the task of sequential recommendation. 
Then we describe the multi-hop time-aware attentive memory recommender overall.

\vspace{-5pt}
\subsection{Preliminaries}

%Sequential recommendation is the task of predicting the next behavior of a user based on the his/her historical behavior sequence.

Suppose there are $M$ users and $N$ items in the system.
The behavior sequence of user $u$ is $\mathbf{S}_u=(b_{u,1}, b_{u,2},...,b_{u,|\mathbf{S}_u|})$.
We denote a behavior $b_{u,i}=(x_{u,i}, t_{u,i}, e_{u,i})$ as the $i$-th interaction in sequence $\mathbf{S}_u$, where $x_{u,i}\in \mathbf{V}$ is the item that user $u$ interacts with at time $t_{u,i}$ and $e_{u,i}$ presents the contextual information.
The contextual information $e_{u,i}$ can include various kinds of important features, e.g. item category, behavior position, location, duration and action.  
%the time-aware sequential recommendation task for addressing the problem, what a user will buy at a target time, is: 
Then given the historical behavior sequence $\mathbf{S}_u=(b_{u,1}, b_{u,2},...,b_{u,i})$ of a specified user $u$, the time-aware sequential recommendation is to predict the next item $x_{u,i+1}$ that user $u$ will interact with at time $t_\text{target}$. 
Since the corpus of item is large in an industrial recommendation system, a nominator in the candidate generation stage needs to make more than one recommendations for the user, which is so-called Top-K recommendation.

We aim to build a time-aware recommender $\mathbf{Rec}$ so that for any prefix behavior sequence $\mathbf{S}_u=(b_{u,1}, b_{u,2},...,b_{u,i})$ and a target time $t_{\text{target}}$, we get the output $y=\mathbf{Rec}(\mathbf{S}_u,t_{\text{target}})$.
As illustrated in Figure \ref{fig:general_framework}, $y=(y_1,y_2,...,y_K)$ is the ranking list for Top-K recommendation ($0<K\ll N$).

\vspace{-5pt}
\subsection{Recommendation with Attentive Memory Network}

User interests are both stable and evolving.
In this paper, we propose a novel attentive memory network for the task of Top-K recommendation, named Multi-hop Time-aware Attentive Memory Network (MTAM for short).
As illustrated in Figure \ref{fig:general_framework}, MTAM first encodes the user's short-term intent into query $q$ via a recurrent network based on our proposed T-GRU and maintains a fixed length history behaviors as long-term preferences in memory $m$.
Then, as shown in Figure \ref{fig:memoryreader}, the prediction procedure is to read the long-term memory $m$ for the current short-term intent $q$ via our proposed time-aware attention mechanism.
The output $a$ of MTAM is a hybrid user representation which takes the advantage of both short-term and long-term components.

The proposed MTAM can be viewed as a non-linear neural generalization of collaborative filtering based on factorization techniques.
Suppose there are $M$ users and $N$ items in a recommendation system.
Let $\mathbf{P}\in \mathbb{R}^{M \times d}$ and $\mathbf{Q}\in \mathbb{R}^{N \times d}$ be the embedding matrices for users and items.
For any prefix behavior sequence $\mathbf{S}_u=(b_{u,1}, b_{u,2},...,b_{u,i})$ of user $u$, we first project items and contextual information into embedding spaces by the look-up function and get $\mathbf{S}_u'=(({b}_{u,1}', t_{u,1}), (b_{u,2}', t_{u,2}),...,(b_{u,i}', t_{u,i}))$.
Our task is building a user model MTAM, of which the output is the embedding of user $u$ at time $t_{\text{target}}$: 
\begin{small} 
 \begin{align}
p_{u,t_{\text{target}}} = \text{MTAM}(\mathbf{S}_u', t_{\text{target}}).
\end{align}
\end{small}
Then, as a neural matrix factorization, a nearest neighbor search can be performed to generate the Top-K recommendations based on the dot-product similarity $p_{u,t_{\text{target}}}\mathbf{Q}^T$ between the predicted user embedding $p_{u,t_{\text{target}}}$ and the embeddings of all items $\mathbf{Q}$:
\begin{small} 
 \begin{align}
y=\mathbf{Rec}(\mathbf{S}_u,t_{\text{target}})=\text{\bf Top-K}(p_{u,t_{\text{target}}}\mathbf{Q}^T),
\end{align}
\end{small}
where $y=(y_1,y_2,...,y_K)$ is the ranking list of $K$ most relevant items.
Our model can be trained by using a standard mini-batch gradient descent on the cross-entropy loss:
\begin{small} 
\begin{align}
L(y',\hat{y})=\sum_{u}^{M} y_u^{'} \log{\hat{y}}_u,
\end{align}
\end{small}
where $y_u^{'}=softmax(p_{u,t_{\text{target}}}\mathbf{Q}^T)$ is the predicted probability distribution of next item and ${\hat{y}}_u$ the one-hot coding of the ground-truth next item $x_{u,i+1}$. 
%Then we will introduce MTAM in detail.

In the following two sections, we first introduce a general methodology to update traditional attentive mechanism and recurrent unit for user modeling, and propose the two basic components of MTAM: time-aware attention mechanism and T-GRU unit in Section 4.
Then we illustrate the proposed multi-hop time-aware attentive memory network which treats the short-term intent as key and reads out the long-term memory attentively for multiple hops in Section 5.

%\section{Time-aware Attention Mechanism and Recurrent Unit}
\section{Temporal Gating Methodology}
In this section, we first give a general idea of temporal gating methodology to capture the time-aware context in user behavior sequences.
Then we describe how to update traditional attention mechanism and recurrent unit with it.

\vspace{-5pt}
\subsection{Temporal Gate}
Different from the semantic correlations between words  in NLP problems, the relationship between two interactions in a behavior sequence is not only related with their relative positions, but also highly influenced by the time intervals.

%[width=2\columnwidth]
%

We model the time interval as a temporal gate to encode the non-linear time difference between two interactions.
A general form of the temporal gate of two behaviors can be defined as:
\begin{small} 
 \begin{equation} 
%\text{Gate} (t_{h_i}, t_{h_j},h_i,h_j) = f(t_{h_i}-t_{h_j},h_i,h_j),
g_{ij}= f(t_{h_i}-t_{h_j},h_i,h_j),
\end{equation}
\end{small} 
where $(h_*,t_{h_*})$ is the hidden representation and timestamp of an interaction. In this way, the temporal relationship between two interactions is determined by the time interval and their respective representations. Next we will introduce how to use the temporal gate to update recurrent networks and attention mechanism.
\vspace{-5pt}
\subsection{Time-aware Recurrent Unit}
Recurrent networks have won great success in user modeling due to their remarkable ability to capture sequential patterns. The recurrent updating function of recurrent networks can be formulated as
\begin{small} 
 \begin{equation} 
h_{s} = f(x_s,h_{s-1}),
\end{equation}
\end{small} 
where $x_s$ is current input and $h_{*}$ is hidden state.
In practice, LSTM and GRU are the two most widely used recurrent units.
The computational complexity of GRU is lower than LSTM by reducing one gate.
In this paper, without loss of generality, we formulate the time-aware recurrent unit with GRU and the computation rules of GRU unit can be illustrated as 
\begin{small} 
 \begin{align} 
z_s &= \sigma([x_s,h_{s-1}]W_z+b_z)\\
r_s &= \sigma([x_s,h_{s-1}]W_r+b_r)\\
h_s^{'} &= \phi ([x_s,h_{s-1}\odot r_s]W_h+b_h)\\
h_s &= z_s \odot h_{s-1} + (1-z_s) \odot h_s^{'},
\end{align}
\end{small}
where $x_s,h_{s-1},h_{s},h_{s}^{'}, z_s, r_s\in \mathbb{R}^{1\times d}$, 
$W_z,W_s,W_h \in \mathbb{R}^{2d \times d}$. 
$[\cdot ,\cdot ]$ is concatenate operation. $+$ and $\odot$ denote element-wise add and multiplication operations.
$\phi$ and $\sigma$ are tanh and sigmoid activation functions.
$z_s$ and $r_s$ are update gate and reset gate, while $h_s^{'}$ is the candidate state, $h_{s-1}$ is history state and $h_s$ is the output hidden state.
The output hidden state $h_s$ emitted at state $s$ is a linear interpolation between the history state $h_{s-1}$ and the candidate state $h_s^{'}$ where update gate $z_s$ acts as a soft switch.

In order to capture the temporal correlations in a user behavior sequence, we design a temporal gate to upgrade GRU, which is jointly determined by the current input $x_s$, the history state $h_{s-1}$ and the time interval $t_s - t_{s-1}$, that is:
\begin{small} 
 \begin{align}
\delta_s &= \phi(\log(t_s - t_{s-1}+1) \odot {W}_{\delta_s} + \mathbf{b}_{\delta_s})\\
\tau_s &= \phi ([x_s,h_{s-1}]W_{\tau_s}+b_{\tau_s})\\
%\text{Gate}(t_x,t_y,x,y) &= \sigma  (\delta\times\mathbf{W}_{\Gamma\delta}+\tau\times\mathbf{W}_{\Gamma\tau}+\mathbf{b}_{\Gamma}),
g_{s}&= \sigma(\delta_s\odot W_{g_{s}\delta_s} +\tau_s\odot W_{g_{s}\tau_s}+b_{g_{s}}),
\end{align}
\end{small} 
where $t_s,t_{s-1}\in \mathbb{R}$, $\delta_s,\tau_s,g_{s}\in \mathbb{R}^{1\times d}$, ${W}_{\delta_s},W_{g_{s}\delta_s},W_{g_{s}\tau_s}\in\mathbb{R}^{1\times d}$, ${b}_{\delta_s},b_{\tau_s},b_{g_{s}}\in\mathbb{R}^{1\times d}$ and ${W}_{\tau_s}\in \mathbb{R}^{2d\times d}$.
The temporal feature $\delta_s$ encodes the temporal relation between adjacent interactions at state $s$.
The semantic feature $\tau_s$ encode the semantic context at state $s$.
Temporal gate $g_s$ is a non-linear combination of temporal and semantic features.

Then, we propose a novel recurrent unit, named T-GRU, by modifying the Eq. (9) to
\begin{small} 
 \begin{align} 
h_s &= z_s \odot g_{s}\odot h_{s-1} + (1-z_s) \odot h_s^{'},
\end{align}
\end{small}
where the temporal gate $g_s$ controls how much past information can be transferred to the current state.
Compared to \cite{yu2019adaptive} which uses two temporal gates to control past and current information, the experimental results show that our proposed T-GRU performs better as a recurrent recommender on 4/6 datasets 
and MTAM dominates on all datasets where T-GRU serves as the short-term intent encoder.

\begin{figure*} 
 \setlength{\belowcaptionskip}{-10pt} 
  \setlength{\abovecaptionskip}{-10pt}   
  \centering
  \includegraphics[width=1.8\columnwidth,height=0.37\textwidth]{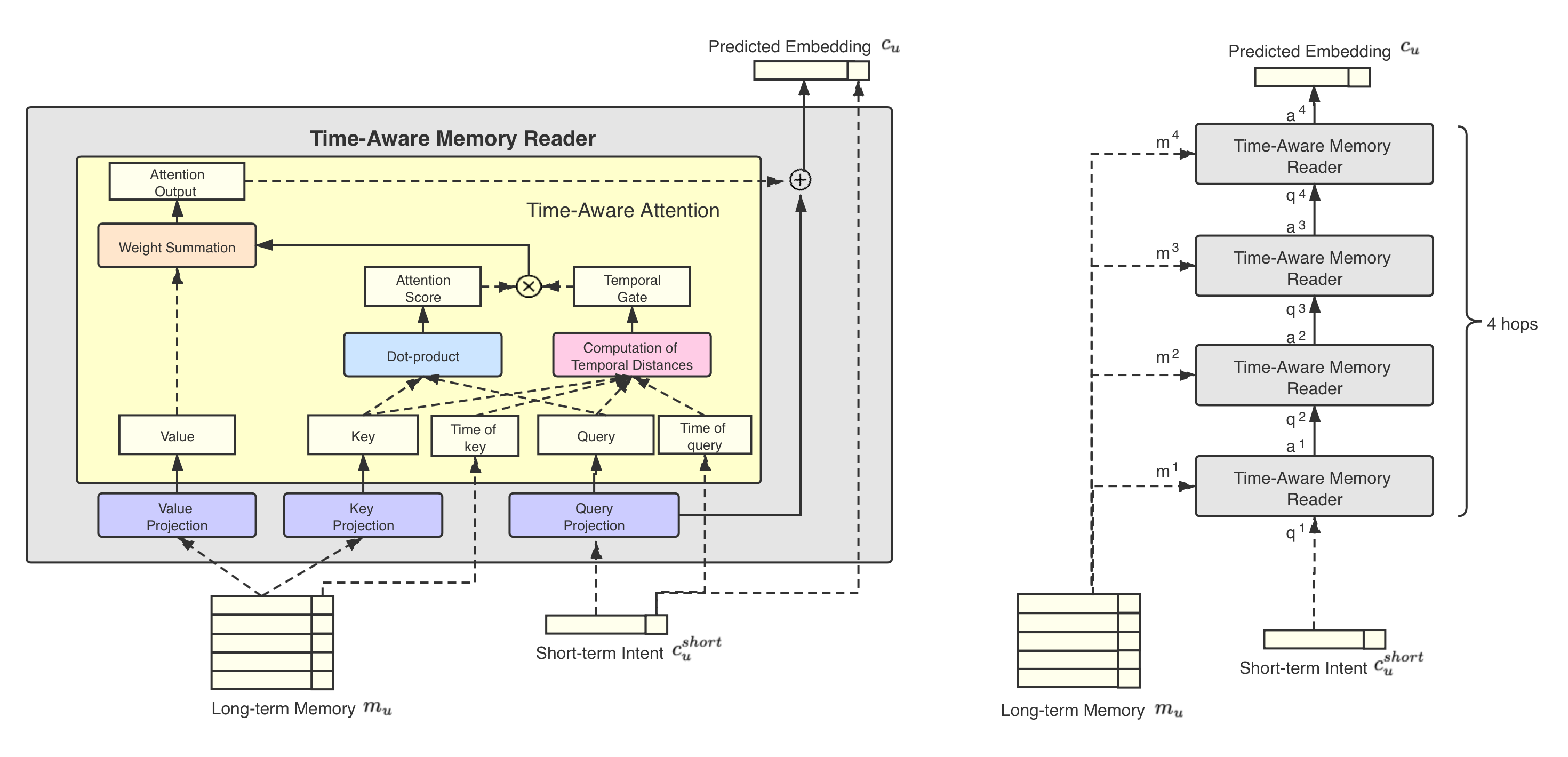}
  \caption{The illustration of multi-hop memory reader, where the left side shows a single layer version and right side shows a four-layer version. Time-aware Memory Reader takes the long-term memory $m_u$ and the current short-term intent $(c_u^{short},t_{target})$ of target user $u$ as input and outputs the current predicted user embedding $(c_u,t_{\text{target}})$. The attention core of MTAM is time-aware attention, where the time-aware attention score combines both temporal and semantic corrections. }
  \label{fig:memoryreader}  
\end{figure*}
\vspace{-5pt}
\subsection{Time-aware Attention}
The scaled dot-product attention \cite{vaswani2017attention} is a popular attention kernel.
Let ${Q}\in\mathbb{R}^{l^Q\times d},{K}\in\mathbb{R}^{l^K\times d},{V}\in\mathbb{R}^{l^V\times d}$ represent query, key and value, where $l^Q, l^K ,l^V$ are the number of items in query, key and value, and $d$ is latent dimension.
%($l^K = l^V$).
The attention correlations between query ${Q}$ and key ${K}$ is computed as 
\begin{small} 
 \begin{equation}
\text{Score}({Q},{K})=\text{softmax}\left ( \frac{{Q}{K}^{T}}{\sqrt{d}} \right ),
\end{equation}
\end{small} 
where the scale factor $\sqrt{d}$ is used to avoid large values of the inner product, especially when the dimension $d$ is high.
$Score({Q},{K}) \in \mathbb{R}^{l_Q\times l_K}$ is a matrix with the shape of the lengths of query ($l_Q$) and key ($l_K$), where the $i$-th row evaluates the relevant percentages of item ${Q_i}$ with all items in key ${K}$.
Then the output of dot-product attention can be computed as a sum of the rows in value ${V}$ weighted by the attention scores, which is formulated as:
\begin{small} 
 \begin{equation} 
\text{Attention}({Q},{K},{V})=\text{score}({Q},{K}){V}.
\end{equation}
\end{small}
\vspace{-10pt}

We propose a time-aware attention mechanism based on scaled dot-product attention, which aims to take the time context into account when calculates the correlation between two items. 
It is a general update of attention mechanism and can be applied in attentive RNNs, self-attention networks and memory networks.
Figure \ref{fig:memoryreader} shows its usage in the memory reader of MTAM.

We first define operation $\hat{-}$ to compute the time interval matrix between behavior sequences. If $A\in \mathbb{R}^m$ and $B\in \mathbb{R}^n$ are two vectors, we define $\hat{-}$ as $C= A \hat{-} B$, such that $C_{ij}=A_i - B_j$ and $C \in \mathbb{R}^{m\times n}$.
For two interaction sequences $(x,t_x)=\{(x_i,t_{xi})\}^{l_x}_i$ and $(y,t_y)=\{(y_i,t_{yi})\}^{l_y}_i$, the temporal gate in time-aware attention can be computed by
\begin{small} 
 \begin{align} 
\delta &= \phi(\log(|t_x \hat{-} t_y|+1) \odot {W}_{\delta} + {b}_{\delta})\\
\tau &= \phi (y {W}_{\tau} x+{b}_{\tau})\\
g_{xy}&= \sigma  (\delta\odot{W}_{g\delta}+\tau\odot{W}_{g\tau}+{b}_{g}),
\end{align}
\end{small}
where $\delta,\tau,g_{xy}\in \mathbb{R}^{l_x\times l_y}$, $x\in \mathbb{R}^{l_x\times d}$, $y\in \mathbb{R}^{l_y\times d}$, $t_x\in \mathbb{R}^{l_x}$, $t_y\in \mathbb{R}^{l_y}$, ${W}_{\tau} \in  \mathbb{R}^{d\times d}$, ${W}_{\delta}, {W}_{g\delta}, {W}_{g\tau}\in  \mathbb{R}^{l_x\times l_y}$ and ${b}_{\delta}, {b}_{\tau}, {b}_{g} \in  \mathbb{R}^{l_x\times l_y}$. 
The temporal feature $\delta$ and semantic feature $\tau$ encode the temporal correlations and semantic correlations between each pair of items in the two sequences.
The temporal gate $g_{xy}\in \mathbb{R}^{l_x\times l_y}$ learns the non-linear correlations between $(x,t_x)$ and $(y,t_y)$ by taking both temporal and semantic information into consideration.
The attention score in Eq. (14) is now changed to:
\begin{small} 
 \begin{equation} 
\text{T-Score}((x,t_x),(y,t_y))=\text{softmax}\left ( \frac{{x}{y}^{T}\odot g_{xy} }{\sqrt{d}} \right ).
\end{equation}
\end{small} 
Finally, we define the time-aware attention of two temporal sequences by updating Eq. (15) as:
\begin{small} 
 \begin{equation} 
\text{T-Attention}(({x},t_x),(y,t_y),(y,t_y))=\text{T-Score}((\mathbf{x},t_x),(y,t_y))y,
\end{equation}
\end{small} 
where the output is a representation of sequence ${x}$ which is temporal related to sequence ${y}$.

\subsection{Discussions}
In this section, we have proposed a methodology to update traditional attentive mechanism and recurrent unit for user modeling.
Despite the detailed differences between temporal gate $g_{xy}$ in dot-product based attention kernel and temporal gate $g_{s}$ in T-GRU unit, $g_{xy}$ and $g_{s}$ are both non-linear functions of time interval between two objects and the semantic contexts of them.
However, intuitively, the preferences tend to be similar within a short period, while large intervals may decrease the influences of the past actions.
We have tried to build the temporal gate as a time-decaying function, for example:
\begin{small} 
 \begin{align}
 g_{ij}=\exp(-\alpha(h_i,h_j)|t_i-t_j|), \text{where } \alpha(h_i,h_j)\geq 0.
\end{align}
\end{small}
But different from non-linear functions $g_{xy}$ and $g_{s}$, it shows no obvious improvement to build a time-decaying temporal gate.
Therefore, unlike our intuition that the correlation between two actions decays with time, the temporal distances are more complex than monotonic decreasing.

\section{Multi-hop Time-aware Attentive Memory Network}
In this section, we describe three components of MTAM in detail. They are short-term intent encoder, long-term memory encoder and reading operation .
\subsection{Short-term Intent Encoder and Long-term Memory Encoder}
%For a behavior sequence $\mathbf{S}_u$ of target user $u$, 
%our memory network takes a memory matrix $m$ and a query $q$, and outputs an answer $a$.
In MTAM, we treat the short-term intent $c_u^{short}$ and a target time $t_{\text{target}}$ as the query $q = (c_u^{short}, t_{\text{target}})$, and store the long-term preference in the memory matrix $m$.
The input of the short-term intent encoder and the long-term memory encoder is a behavior sequence $\mathbf{S}_u'=(({b}_{u,1}', t_{u,1}), (b_{u,2}', t_{u,2}),...,(b_{u,i}', t_{u,i}))$, where the $i$-th behavior $(b_{u,i}', t_{u,i})$ is a tuple of behavior embedding $b_{u,i}'\in\mathbb{R}^{1\times d}$ and timestamp $ t_{u,i}\in\mathbb{R}$.
\vspace{0pt}

\subsubsection{Short-term Intent Encoder}
In the short-term intent encoder, we build a recurrent network with T-GRU unit (introduced in Section 4.3) rather than traditional GRU \cite{li2017neural} and LSTM \cite{li2018learning} units to encode the current intents of users.
T-GRU filters out irrelevant historical information by controlling how much past information can be transferred to future states via the temporal gate (Eq. (12)).
Therefore, the proposed T-GRU unit is more capable than traditional and existing time-aware recurrent units \cite{zhu2017next,yu2019adaptive} to capture the current intents of users.
We use the final hidden state $h_{u,i}$ as the short-term intent representation of user $u$: $c_u^{\text{short}}= h_{u,i}$,
where $h_{u,i} \in \mathbb{R}^{1\times d}$ is the short-term intent representation.

%\vspace{-5pt}
\subsubsection{Long-term Memory Encoder}
Long-term memory encoder, also can be called long-term memory writer, maintains user's prior records in a personalized memory $m$.
Memory $m = (m_1,m_2,...,m_L)$ is a fixed-length queue with $L$ slots and each memory slot $m_i \in m$ stores a user historical record $(b_{u,*}', t_{u,*})$. 
As mentioned in many researches \cite{chen2018sequential,liu2018stamp}, users' recent behaviors usually are more important to the current predictions. 
We adopt a simple first-in-first-out rule and maintain the latest $L$ behaviors of $\mathbf{S'}_u$ in the user memory $m$, which is 
\vspace{-5pt}
\begin{small} 
\begin{align}
m_u =((b_{u,i}', t_{u,i}),(b_{u,i-1}', t_{u,i-1}),...,(b_{u,i-L+1}', t_{u,i-L+1})).
%where m_{u,b} = (b_{u,i}',b_{u,i-1}',...,b_{u,i-L+1}'),\\
%m_{u,t} = (t_{u,i}',t_{u,i-1}',...,t_{u,i-L+1}'),
\end{align}
\end{small}
In the experiments, we empirically set $L$ as same as the maximum length of $\mathbf{S'}_u$.
If length of $\mathbf{S'}_u$ is less than $L$, we add zero-paddings to the right side of $m_u$ to convert the $m_u$ to a fixed-length queue.
\vspace{-5pt}
\subsection{Reading Operation}
The reading operation is the key component of a memory network, which determines how to predict the answer based on a query and the information that stored in the memory.

 \vspace{-5pt}
\subsubsection{Single Layer}
We start by describing the memory reader of MTAM in the single layer case, then show how to stack it for multiple hops.

The memory reader of MTAM reads the memory $m_u$ attentively for a given query $(c_u^{\text{short}}, t_{\text{target}})$ and outputs a predicted user embedding $c_u$. 
We use the time-aware attention which is introduced in Section 4.3 as the attention kernel.
As illustrated in Figure \ref{fig:memoryreader}, the time-aware memory reader projects the current intent $c_u^{short}$ and the behavior embeddings in memory $m_{u,b}$ into value, key and query, and computes the attention output as:
\begin{small} 
\begin{align}
o_u = \text{T-Attention}((c_u^{\text{short}}W_Q,t_{\text{target}}),(m_{u,b}{W}_K,t_{u,t}),(m_{u,b}{W}_V,t_{u,t})),
\vspace{-1cm}
\end{align}
\end{small}
where $W_Q,W_K,{W}_V\in\mathbb{R}^{d\times d}$ are query weight, key weight and value weight, $m_{u,b}=(b_{u,i}',b_{u,i-1}',...,b_{u,i-L+1}')$ is the behavior memory and $t_{u,t}=(t_{u,i}, t_{u,i-1},..., t_{u,i-L+1})$ is the time memory.

Then we compute the output of memory reader, which is the predicted embedding of user $u$ at time $t_{\text{target}}$, as: $c_u = c_u^{\text{short}}+ o_u$.
%\begin{small} 
%\begin{align}
%c_u = c_u^{\text{short}}+ o_u.
%\end{align}
%\end{small}
%\vspace{-5pt}

\subsubsection{Multiple Layers}
Inspired by previous works \cite{sukhbaatar2015end,tran2019signed} where the multi-hop designs improve the performance of memory networks, we stack the single layer memory readers to construct a deeper network (MTAM).
We illustrate how to build a multi-hop memory reader in Figure \ref{fig:memoryreader}.
Let the short-term intent be the query for the first hop. 
Then, the multi-hop memory reader can be formulated recurrently, where the output of the $k$-th hop is:
\begin{small} 
\begin{align}
o_u^k &= \text{T-Attention}((c_u^{k-1}{W}_Q^k,t_{\text{target}}),(m_{u,b}{W}_K,t_{u,t}),(m_{u,b}{W}_V,t_{u,t}))\\
c_u^k &= c_u^{k-1}+ o_u^k.
\end{align}
\end{small}
Performing the reading operation for multiple hops helps us to capture the diversity of user preference, because the memory reader in different hops may concentrate on different behaviors.
For a MTAM network with $k'$ hops, the output, which is the predicted user embedding in Eq. (1), is $p_{u,t_{\text{target}}} =\text{MTAM}({S_u}',t_{\text{target}})= c_u^{k'}$.

\vspace{-5pt}
\section{Experiment}
In this section, we first describe the setups of all experiments.
Then, we demonstrate the effectiveness of our proposed models from the following aspects:
(1) The performance of the proposed framework and comparable methods.
(2) The effectiveness of the proposed time-aware attention kernel, T-GRU unit and the multi-hop structure of MTAM.
(3) The Influence of Multiple Hops.
\subsection{Datasets}
We conduct experiments on six real-world datasets of two types. MovieLens-20 and Amazon datasets are rating datasets, which are not "real" user behavior logs, but consist of comments on items. Yoochoose and Ali Mobile are transaction datasets, which directly record the behavior trajectories of users.

\begin{table}[h]\small
    \setlength{\belowcaptionskip}{0pt} 
    \setlength{\abovecaptionskip}{0pt}
  \centering
  \caption{The Statistics of Datasets}
  \label{tab:Statistics}  
  \resizebox{0.45\textwidth}{14.5mm}{
    \begin{tabular}{c|ccccc}
      \toprule[1pt]
    Statistics & \#user  & \#Item & \#Cat. &\makecell[c]{Avg. behaviors \\ per user} &Density\\   
    \toprule[1pt]
    ml-25m &  12015  &   4991 & 712& 104.27& 2.0891\%\\
    Electronics & 41940 & 87203 &1063 &31.67& 0.0363\%\\
    CDs \& Vinyl & 39663 & 33593 &  419 & 24.78& 0.0738\%\\
    Movies \& TV & 105321 &35164&379&22.70&0.0645\%\\
    Yoochoose 1/4& 108817&16296&187&15.52&0.0953\%\\
    Ali Mobile & 9980 & 594083&6352&1226.92& 0.1267\%\\
       \toprule[1pt] 
    \end{tabular}}%
    \vspace{-15pt}
\end{table}%
\begin{itemize}
\item \textbf{MovieLens}\footnote{https://grouplens.org/datasets/movielens/20m/} is a widely used benchmark dataset for evaluating collaborative filtering algorithms. We use the latest stable version (MovieLens-25m) which includes 25 million user ratings. 

\item \textbf{Amazon}\footnote{http://deepyeti.ucsd.edu/jianmo/amazon/index.html} is a popular dataset to evaluate recommendation algorithms.
It is always used as a benchmark for sequential recommendation tasks.
We consider three categories: Electronics, CDs \& Vinyl and Movies \& TV. 

\item \textbf{Yoochoose}\footnote{http://2015.recsyschallenge.com/challenge.html} from the RecSys'15 Challenge I contains click-streams gathered from an e-commerce web site in six months. Because the Yoochoose dataset is quite large, we randomly sample 1/4 users.

\item \textbf{Ali Mobile}\footnote{https://tianchi.aliyun.com/dataset/dataDetail?dataId=46} from the Alibaba Competition contains transaction data gathered from Alibaba's M-Commerce platform in one month.
\end{itemize}

We filter the users whose behavior lengths are less than 10 and items that appear less than 30 times.
The statistics of six datasets after data preprocessing are shown in Table \ref{tab:Statistics}.
Although there are various types of context information in these datasets (e.g. actions, comments, descriptions), We only use the category of an item and the position of a behavior in a sequence as context features in our experiments.
For a behavior sequence $\mathbf{S}_u=(b_{u,1}, b_{u,2},...,b_{u,i})$, we use a sequence splitting preprocess to generate the sequences and corresponding labels $([b_{u,1}], b_{u,2}), ([b_{u,1}, b_{u,2}], b_{u,3}),...,([b_{u,1}, b_{u,2} , ...,\\b_{u,i-2}], b_{u,i-1})$ for the training set and the last behavior $([b_{u,1}, b_{u,2} , ...,\\b_{u,i-1}], b_{u,i})$ for the testing set.

\begin{table*}[!tb]\small
    \setlength{\belowcaptionskip}{0pt} 
    \setlength{\abovecaptionskip}{0pt}
  \centering
  \caption{Performance comparison of MTAM and the baseline methods. Among these baselines, \textbf{T-GRU} is a component of MTAM, and \textbf{NARM+} and \textbf{NARM+} is implemented with our proposed T-GRU and time-aware attention.
 We divide models into 4 groups: naive recommenders (e.g. Top Pop), non-hybrid recommenders (e.g. GRU), hybrid recommenders (e.g. NARM) and the proposed MTAM.
%We also divide datasets into 2 groups: rating datasets (e.g. ML-20) and transaction datasets (e.g. Yoochoose). 
The underlined number is the best baseline method and the boldfaced number is the best method of all. Improv. denotes the improvement of the best model over the best baseline method.
Significant differences are with respect to the best baseline methods.
}
  \label{tab:performance}  
  \resizebox{\textwidth}{27.5mm}{
	\begin{tabular}{c|ccc|cccccc|cccc|ccc}
		\toprule[1pt]
		~ &\multicolumn{16}{c}{HR@10 } \\
		\cline{2-17}
		~ & Top Pop&P-Pop & BPR-MF  & GRU$-$$-$ & GRU & T-SeqRec  & \textbf{T-GRU} &SASRec & TiSASRec & NARM  &  \textbf{NARM+} &  \textbf{NARM++} &STAMP& \textbf{MTAM}& improv. & p-value\\
		\toprule[1pt]
		ML-25m& 0.0300&0 & 0.0108  &0.1985& 0.1946 &  \underline{\textit{0.2001}}& 0.2006 &0.1847 &0.1865 &0.1999&0.1978&0.1981&0.1821&\textbf{0.2053}&2.60\% &3.39e-3\\
		\hline
		Electronics &0.0108& 0.0017 & 0.0204 &0.0283& 0.0370& \underline{\textit{0.0376}}& 0.0384 &0.0328 &  0.0332 &0.0334&0.0322&0.0371&0.0330&\textbf{0.0423}&12.5\%&6.12e-11 \\
		\hline
		CDs \& Vinyl & 0.0062&0.0028 & 0.0330  &0.1070& 0.1082& 0.1130& 0.1129 &0.1053 &0.1075 &\underline{\textit{0.1131}}&0.1111&0.1160&0.1091&\textbf{0.1206}&6.63\%&6.65e-3 \\
		\hline
		Movies \& TV & 0.0114&0.0021 & 0.0279 &0.1512&0.1505& \underline{\textit{0.1550}}& 0.1523 & 0.1422 &0.1441 &0.1499&0.1548&0.1583&0.1372&\textbf{0.1605}&3.54\%&6.96e-5\\
		\toprule[1pt]
		Yoochoose &0.0212& 0.2599& 0.3263 & 0.5273 & 0.5345 &0.5332& 0.5351 &0.5240 &0.5270  &0.5356&0.5355&0.5360&\underline{\textit{0.5360}}&\textbf{0.5386}&0.49\%&5.40e-4\\
		\hline
		Ali Mobile & 0.0046 &0.1366 & 0.1533 &0.2321& 0.2299& \underline{\textit{0.2380}}&  0.2427 &0.2274 &0.2292 &0.2318&0.2329&0.2427&0.2097&\textbf{0.2501}&5.08\%&7.99e-4\\
		\toprule[1pt]	
		~ &\multicolumn{16}{c}{NDCG@10 }  \\
		\toprule[1pt]
		ML-25m&  0.0140&0 & 0.0050 &0.1146&  0.1112 & \underline{\textit{0.1159}} & 0.1150 &0.1001 &0.1002 &0.1129&0.1127&0.1121&0.0981& \textbf{0.1187}&2.42\%&6.52e-3\\
		\hline
		Electronics & 0.0051&0.0009 & 0.0119 &0.0180&\underline{\textit{0.0267}}& 0.0257& 0.0263 &0.0229 & 0.0231&0.0221&0.0205&0.0249&0.0232&\textbf{0.0307}&15\%&3.81e-9\\
		\hline
		CDs \& Vinyl & 0.0030&0.0016 & 0.0138 &0.0715& 0.0705& \underline{\textit{0.0756}}& 0.0754 &0.0667 &0.0679 &0.0747&0.0724&0.0746&0.0723&\textbf{0.0783}&3.57\%&6.51e-3 \\
		\hline
		Movies \& TV  & 0.0056&0.0013 & 0.0135 &0.1135&  0.1129& \underline{\textit{0.1167}}& 0.1138 & 0.1045 &0.1061 &0.1116&0.1168&0.1169&0.1035&\textbf{0.1210}&3.68\%&4.37e-5\\
		\toprule[1pt]
		Yoochoose &  0.0102&0.1698 & 0.1977& 0.3313 & 0.3365 &0.3360& 0.3363 &0.3293 &0.3295 &0.3372&0.3384&\textbf{0.3386}&\underline{\textit{0.3380}}&0.3381&0.12\% & 6.35e-4\\
		\hline
		Ali Mobile & 0.0023&0.0834 & 0.1011 &0.1562& 0.1549& \underline{\textit{0.1602}}& 0.1615 &0.1524 &0.1541 &0.1555&0.1560&0.1621&0.1424&\textbf{0.1651}&3.06\%&3.83e-3\\
		\toprule[1pt]

	\end{tabular}
}
\end{table*}%

\subsection{Compared Methods and Implementation Details}
We compare \textbf{MTAM} with the following competitive models:
\begin{itemize}
\item \textbf{Top Pop/P-Pop} recommends items of the largest interactions with all users/a target user. They are commonly used baselines for all recommendation researches.
\item \textbf{BPR-MF} \cite{rendle2009bpr} is matrix factorization recommender for the personalized ranking task.
\item \textbf{GRU} is a classical recurrent sequential recommender.
\item \textbf{GRU$-$$-$} is \textbf{GRU} without the category features.
\item \textbf{T-SeqRec} equips LSTM with two temporal gates to model time intervals and time spans for recommendation problem \cite{yu2019adaptive}. Since it dominates other time-aware RNN-based models (e.g. \cite{zhu2017next,chen2019dynamic}), we treat it as a state-of-art time-aware recurrent unit.
 %It is proposed in \cite{yu2019adaptive} to capture both long-term and short-term interests.

\item \textbf{T-GRU} is proposed in this paper, which equips GRU with a new temporal gate. Different from \textbf{T-SeqRec}, \textbf{T-GRU} only focus on short-term preference without using time spans to capture long-term preference.
\item \textbf{SASRec} \cite{kang2018self} is a self-attention based sequential recommender.
\item \textbf{TiSASRec} \cite{li2020time} is a time-aware self-attentive sequential recommender which directly adds the clipped relative time intervals to the dot-product of item embeddings.
\item \textbf{NARM} \cite{li2017neural} is a hybrid sequential recommender which utilizes attention mechanism to model the user's local purposes. It is a commonly used baseline hybrid recommender which takes both local and global preferences into consideration.

\item \textbf{NARM+} is improved by equipping with our proposed time-aware attention.

\item \textbf{NARM++} is improved by equipping with our proposed time-aware attention and T-GRU.

%\item \textbf{LSTUR} \cite{an2019neural} combines long-term and short-term preferences by using the long-term representation to initialize the hidden state of the short-term GRU network.

\item \textbf{STAMP} \cite{liu2018stamp} captures both long-term and short-term preferences using an attentive MLP network.
\end{itemize}

To ensure fair comparison, we set all hidden units and low-rank embedding spaces, including RNN layers and attention layers, as 128.
We set the initial learning rate as 1e-3 and use an exponential learning rate decay for every 100 iterators with 0.995 decay rate. 0.5 dropout rate and 1e-5 regularization rate are used to reduce overfitting. The maximum length of user behavior is set to 50.
The proposed models and all compared models are implemented with tensorflow 1.14\footnote{https://github.com/cocoandpudding/MTAMRecommender}, and trained and tested on a linux server with a Tesla P100 GPU.

\vspace{-0.3em}

\subsection{Evaluation Metrics}
Since items that an individual can interact with are extremely sparse, recommenders can suggest a set of candidate items each time.
We use HR@k and NDCG@k as the metrics for all models, where $k$ is the number of items recommended each time.

\noindent\textbf{HR@k} is short for Hit Ratio, which shows whether the target item is in the recommended list or not. Since we only consider one ground truth for each sample, HR@k is equivalent to Recall@k.

\noindent\textbf{NDCG@k} takes the position of hit item into account by assigning a highest score to the hit at top rank and decreasing the scores to hits at lower ranks. NDCG, short for Normalized Discounted Cumulative Gain, not only considers the HR but also the orders of ranking.

Statistical significance of observed differences between the performance of the purposed MTAM and the best baseline methods is tested by the t-test on pair-wise samples. 
Small p-values are associated with large t-statistics, where the threshold 0.01 means strong significance and the threshold 0.05 means weak significance.

\subsection{Overall Recommendation Performance}

\begin{figure}[th]
    \setlength{\belowcaptionskip}{-15pt}
    \setlength{\abovecaptionskip}{-5pt}
    \centering
    \subfigure[Recall on Amazon Electronics]{
        \includegraphics[width=0.225\textwidth, height=0.142 \textwidth]{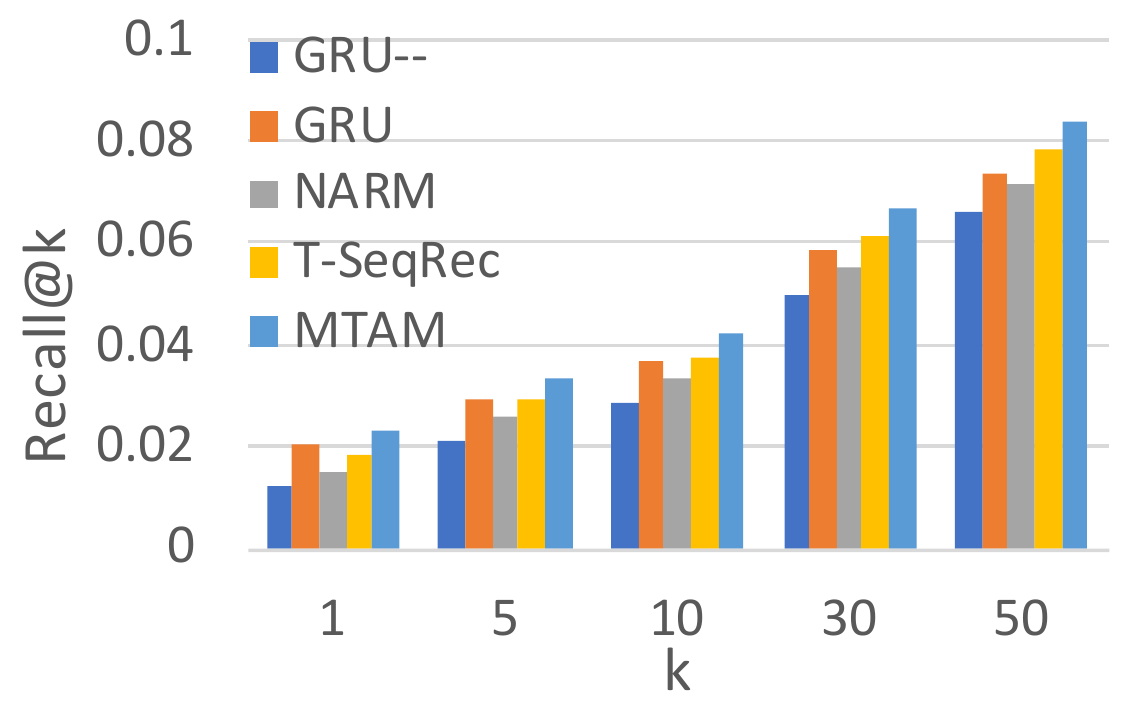}
    }
     \hspace{0pt}
        \subfigure[NDCG on Amazon Electronics]{
        \includegraphics[width=0.225\textwidth, height=0.142\textwidth]{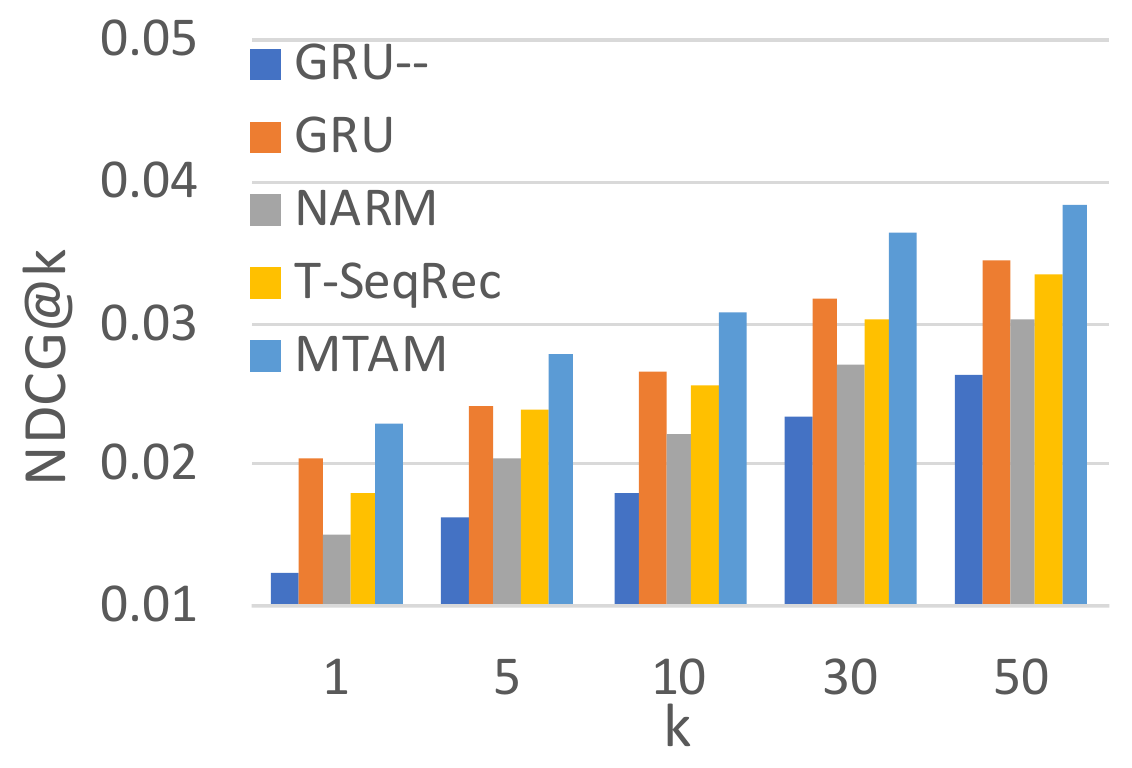}
    }
    \caption{The overall performance comparison on Amazon Electronics.}
    \label{fig:Electronics_k}
\vspace{-10pt}
\end{figure}

\begin{figure}[t]
 % \vspace{-15pt}
    \setlength{\belowcaptionskip}{-15pt}
    \setlength{\abovecaptionskip}{-5pt}
    \centering
    \subfigure[HR on Ali Mobile]{
        \includegraphics[width=0.225\textwidth, height=0.14\textwidth]{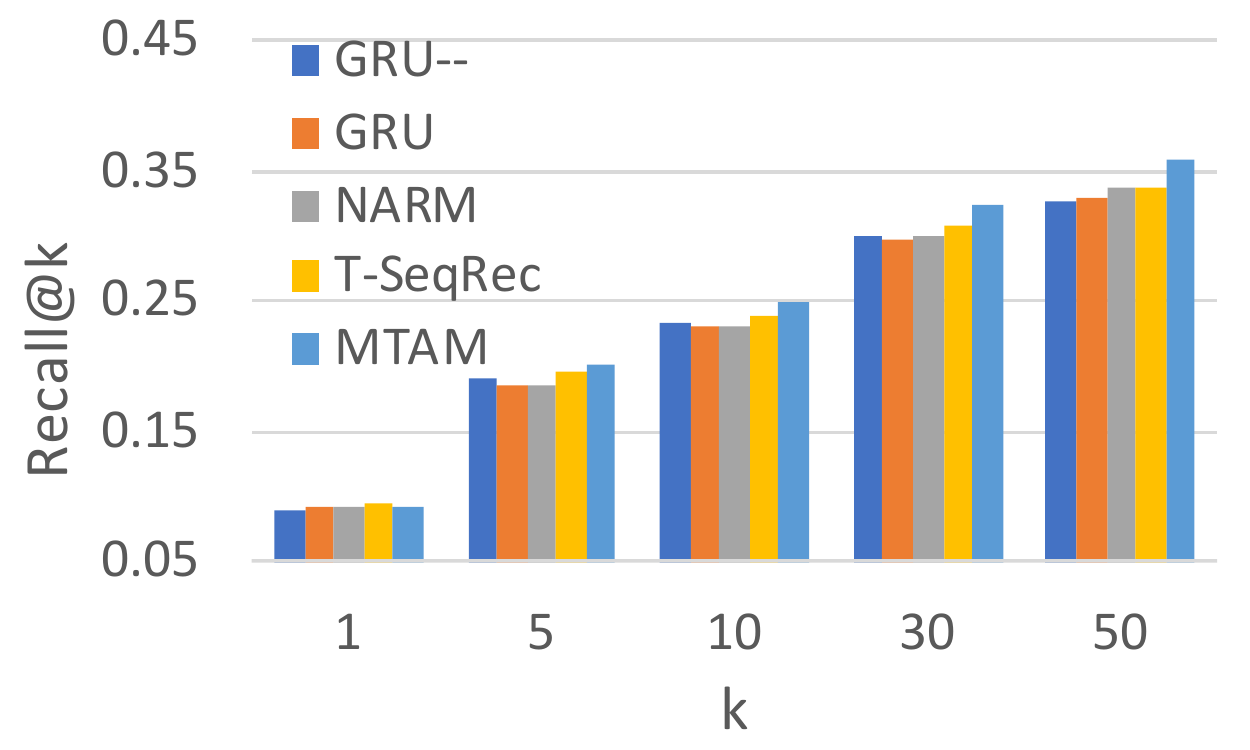}
    }
     \hspace{0pt}
        \subfigure[NDCG on Ali Mobile]{
        \includegraphics[width=0.225\textwidth, height=0.14\textwidth]{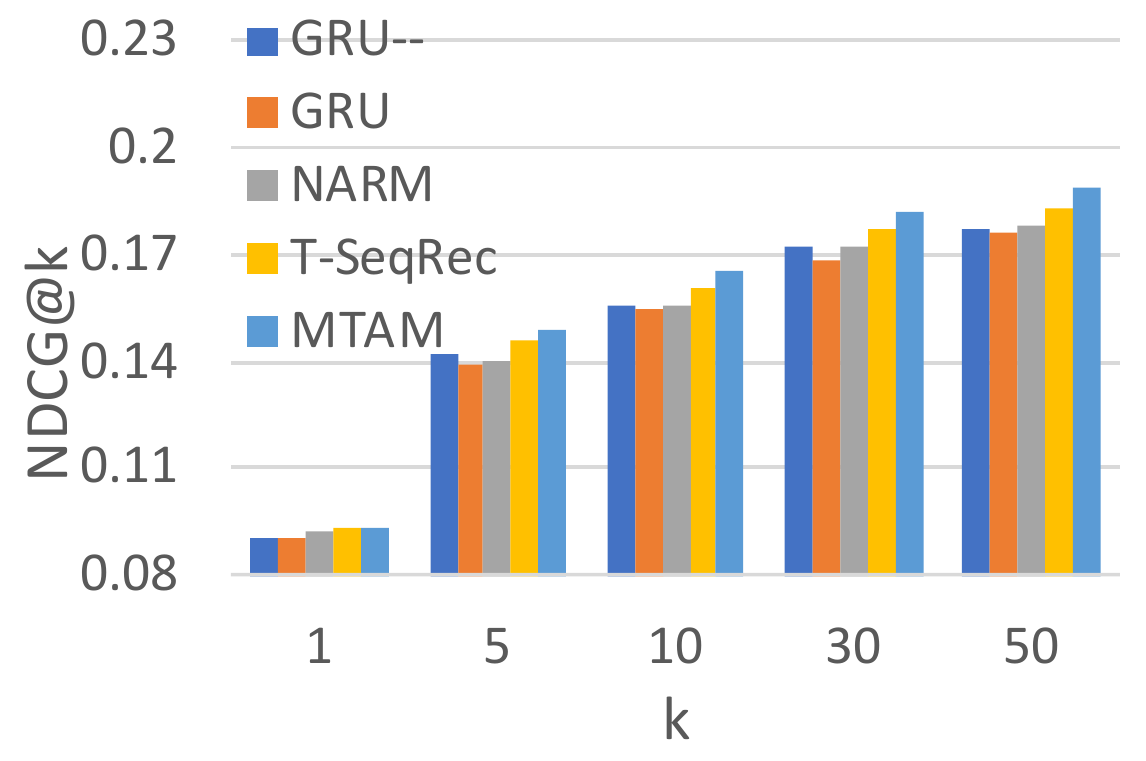}
    }
    \caption{The overall performance comparison on Ali Mobile.}
    \label{fig:Ali_k}
\end{figure}

Table~\ref{tab:performance} and Figure~\ref{fig:Electronics_k}, \ref{fig:Ali_k} illustrate the performance of \textbf{MTAM} and the baseline methods. 
%As the results of different $K$s are consistent for most cases, 
We only report the results of $K = 10$ for all models and all datasets in Table~\ref{tab:performance} due to space limitation.
And in Figure~\ref{fig:Electronics_k}, \ref{fig:Ali_k} , we take the one rating dataset (Electronics) and one transaction dataset (Ali Mobile) as representatives to show the results of different $K$s for typical models.

%\begin{figure*}
   % \setlength{\belowcaptionskip}{-15pt} 
  %  \setlength{\abovecaptionskip}{-5pt}
  %  \centering
  %  \subfigure[Recall on Amazon Electronics]{
 %       \setlength{\belowcaptionskip}{-10pt} 
 %   \setlength{\abovecaptionskip}{-50pt}
   %     \includegraphics[width=0.42\textwidth,height=0.148\textwidth]{elec_recall_new.pdf}
  %  }
  %  \hspace{0pt}
   %     \subfigure[NDCG on Amazon Electronics]{
   %         \setlength{\belowcaptionskip}{-10pt} 
  %  \setlength{\abovecaptionskip}{-50pt}
 %       \includegraphics[width=0.42\textwidth,height=0.148\textwidth]{elec_ndcg_new.pdf}
%    }
%    \hspace{0pt}
%    \caption{The performance comparison of ablation analysis on Amazon Electronics.}
%    \label{fig:AblationElectronics} 
%\end{figure*} 

%\begin{figure*}
%    \setlength{\belowcaptionskip}{-15pt} 
%    \setlength{\abovecaptionskip}{-5pt}
%    \centering
%    \subfigure[Recall on Ali Mobile]{
%        \includegraphics[width=0.42\textwidth,height=0.18\textwidth]{taobaoapp_recall_new.pdf}
%    }
%    \hspace{0pt}
%        \subfigure[NDCG on Ali Mobile]{
%        \includegraphics[width=0.42\textwidth,height=0.18\textwidth]{taobaoapp_ndcg_new.pdf}
%    }
    %\hspace{0pt}
%    \caption{The performance comparison of ablation analysis on Ali Mobile.}
%    \label{fig:AblationAli} 
%\end{figure*}

In general, the proposed MTAM outperforms the state-of-the-art methods significantly (p-value<0.01).
Our \textbf{MTAM} has achieved the best performances on five datasets expect Yoochoose.
On Yoochoose, \textbf{MTAM} performs the best for HR@10, but is defeated by \textbf{NARM++} for NDCG@10.
\textbf{NARM++} is an attentive RNN model which is updated by the proposed time-aware attention and T-GRU.
%Therefore, the overall experimental results demonstrate the effectiveness of our proposed methods.
Therefore, the overall experimental results demonstrate the effectiveness of our proposed methods.
Comparing the results, we have four observations:

(1) Compared with the models based on traditional RNN and attention mechanism, time-aware neural networks perform obviously better on all datasets. 
%One possible explanation is that short-term intents influence users's purchase decisions stronger than long-term preference in online shopping scenario.
For example, \textbf{T-SeqRec} dominates all baseline models in four of the six datasets, where it even achieves much better performance than \textbf{NARM}.
And \textbf{NARM+} outperform \textbf{NARM} on three datasets, while \textbf{NARM++} outperform both \textbf{NARM+} and \textbf{NARM} on most all dataset (\textbf{NARM++} has comparable performance to \textbf{NARM+} on ML-25m dataset).
These results confirm that temporal information contributes to recommendation performance  obviously and further demonstrate that the proposed time-aware attention and T-GRU help a lot to build user models better.

(2) When talking about whether hybrid recommenders that considers both long-term and short-term preferences provide more competitive results, it is a little complicated.
First, we observe that \textbf{MTAM} dominates all pure RNN-based or self-attention based models, but \textbf{NARM} loses to \textbf{GRU} on two datasets.
 Then, we can see that \textbf{NARM++} performs better than all pure models.
These results indicate that leveraging both long-term and short-term preferences leads to substantial performance improvement in sequential recommendation, but the hybrid models should have the ability to model the temporal dynamics, diversity and complexity of users' sequential behaviors.
The outperformance of \textbf{MTAM} demonstrates that the proposed multi-hop memory network is good at dealing with sequential information with the help of time-aware attention and T-GRU.

(3) The results of \textbf{Top Pop} and \textbf{P-Pop} are quite different on rating data and transaction data. We observe that the simple \textbf{Top Pop} provides not bad baseline results on ML-25m and all Amazon datasets, but it hardly works on Yoochoose and Ali Mobile datasets.
On the other hand, \textbf{P-Pop} performs well on Yoochoose and Ali Mobile datasets, but doesn't work on the rating datasets.
These results directly reveal the different patterns in rating sequences and transaction sequences.
Users may not rate on the same item multiple times in rating cases, which explains the poor performance of \textbf{P-Pop} on rating datasets.
While in transaction scenarios, users tend to click on the same item many times and the more interactions may indicate more interest, which explains the good performance of \textbf{P-Pop} on transaction datasets.
The experimental results demonstrate that \textbf{MTAM} performs well on both rating data and transaction data.

\begin{table*}[h]\large
    \setlength{\belowcaptionskip}{-5pt} 
    \setlength{\abovecaptionskip}{0pt}
  \centering
  \caption{Ablation Analysis on six datasets. MTAM with T-SeqRec uses T-SeqRec as the recurrent unit. MTAM without Time-Aware RNN uses traditional GRU as recurrent unit and MTAM without Time-Aware attention uses traditional dot-product attention as attention kernel. MTAM via T-GRU and MTAM via GRU store the hidden states of RNN instead of behavior embeddings. Time-Aware Self Attention is a SASRec implemented with the proposed time-aware attention. We treat GRU as the baseline model and the boldface number is the best method. We show the improvement of the best method over GRU.}
  \label{tab:Ablation}
  \resizebox{\textwidth}{24mm}{  
	\begin{tabular}{c|cccccc|cccccc}
		\toprule[1pt]
		~ &\multicolumn{6}{c}{HR@10 }& \multicolumn{6}{c}{NDCG@10 }   \\
		\cline{2-13}
		~ & ML-25m & Electronics &CDs \& Vinyl  &  Movies \& TV  & Yoochoose & Ali Mobile& ML-25m & Electronics &CDs \& Vinyl  &  Movies \& TV  & Yoochoose & Ali Mobile\\
		\toprule[1pt]
		GRU &  \underline{\textit{0.1946}} & \underline{\textit{0.0370}} &  \underline{\textit{0.1082}} & \underline{\textit{0.1474}}&\underline{\textit{0.5345 }}&\underline{\textit{0.2299 }} & \underline{\textit{0.1159}}&\underline{\textit{0.0267}}&\underline{\textit{0.0705}}&\underline{\textit{0.1104}}&\underline{\textit{0.3365}}&\underline{\textit{0.1549}} \\
		\hline
		MTAM& 0.2053& \textbf{0.0423} &  \textbf{0.1206} & 0.1605&0.5386 &0.2501 &0.1187& \textbf{0.0307} &\textbf{0.0783}&0.1210& 0.3381&0.1651 \\
		\hline
		\makecell[c]{MTAM with\\ T-SeqRec} &\textbf{0.2073(1.0\% $\uparrow$)}&0.0396(6.4\% $\downarrow$)&0.12(0.5\% $\downarrow$)&0.1603(0.1\% $\downarrow$)&0.5378(0.2\% $\downarrow$)&0.2531(1.2\% $\uparrow$)&\textbf{0.1203(1.3\% $\uparrow$)}&0.0289(5.9\% $\downarrow$)&0.0789(0.8\% $\uparrow$)&0.1213(0.3\% $\uparrow$)&0.337(0.3\% $\downarrow$)&0.1671(1.2\% $\uparrow$)\\
		\hline
		\makecell[c]{MTAM without \\ Time-Aware \\RNN} &0.2038(0.7\% $\downarrow$)&0.025(41\% $\downarrow$)&0.1135(5.9\% $\downarrow$)&0.1603(0.1\% $\downarrow$)&0.5389(0.1\% $\uparrow$)&0.2519(0.7\% $\uparrow$)&0.1171(1.3\% $\downarrow$)&0.0156(49\% $\downarrow$)&0.075(4.2\% $\downarrow$)&0.1212(0.1\% $\uparrow$)&0.3378(0.1\% $\downarrow$)&0.1669(1.0\% $\uparrow$)\\

		\hline
		\makecell[c]{MTAM without \\ Time-Aware\\ Attention} & 0.1952(4.9\%$\downarrow$) & 0.0375(11\%$\downarrow$) &0.1127(6.6\%$\downarrow$) &  \textbf{0.1612}(0.4\%$\uparrow$)   &0.5361(0.4\%$\downarrow$) &0.2420(3.2\%$\uparrow$) &0.1116(6.0\%$\downarrow$) & 0.0264(14\%$\downarrow$)  & 0.0727(7.2\%$\downarrow$) &\textbf{0.1218}(0.7\%$\uparrow$)&0.3367(0.4\%$\downarrow$)&0.1603(2.9\%$\downarrow$)\\
		\hline
		MTAM via T-GRU &0.2032(1.0\% $\downarrow$) &0.0369(13\% $\downarrow$)&0.1136(5.8\% $\downarrow$)&0.1611(0.4\% $\uparrow$)&0.5348(0.7\% $\downarrow$)&0.25161(0.6\% $\uparrow$)&0.1161(2.2\% $\downarrow$)&0.0246(20\% $\downarrow$)&0.0738(5.8\% $\downarrow$)&0.1205(0.4\% $\downarrow$)&0.3376(0.2\% $\downarrow$)&0.1673(1.3\% $\uparrow$)\\

		\hline
		MTAM via GRU &0.2002(2.5\% $\downarrow$)&0.0255(40\% $\downarrow$)&0.1095(9.2\% $\downarrow$)&0.1593(0.8\% $\downarrow$)&0.5352(0.6\% $\downarrow$)&0.2501(-)&0.1141(3.8\% $\downarrow$)&0.0154(50\% $\downarrow$)&0.0717(8.4\% $\downarrow$)&0.1207(0.3\% $\downarrow$)&0.3366(0.4\% $\downarrow$)&0.1652(0.1\% $\uparrow$)\\

		\hline
		\makecell[c]{Time-Aware \\Self Attention} &0.1953(4.7\% $\downarrow$)&0.038(10\% $\downarrow$)&0.1112(7.8\% $\downarrow$)&0.1566(2.4\% $\downarrow$)&\textbf{0.5393}(0.1\% $\uparrow$)&\textbf{0.2549}(1.9\% $\uparrow$)&0.1112(6.3\% $\downarrow$)&0.0283(7.8\% $\downarrow$)&0.0736(6.0\% $\downarrow$)&0.1175(2.9\% $\downarrow$)&\textbf{0.3394}(0.4\% $\uparrow$)&\textbf{0.1682}(1.9\% $\uparrow$)\\
		\toprule[1pt]
		improvement & 6.5\%& 14.3\% & 11.5\% &9.4\%&0.90\%&8.8\% &3.8\%&15.0\%&11.1\%&10.3\%&0.86\%&8.6\% \\
		\toprule[1pt]
		%\vspace{-5pt}
	\end{tabular}}
\end{table*}%

\begin{figure*}
\vspace{-15pt}
    \setlength{\belowcaptionskip}{-15pt} 
    \setlength{\abovecaptionskip}{-5pt}
    \centering
    \subfigure[HR@10 and NDCG@10 on ML-25m]{
        \includegraphics[width=0.1545\textwidth, height=0.13\textwidth]{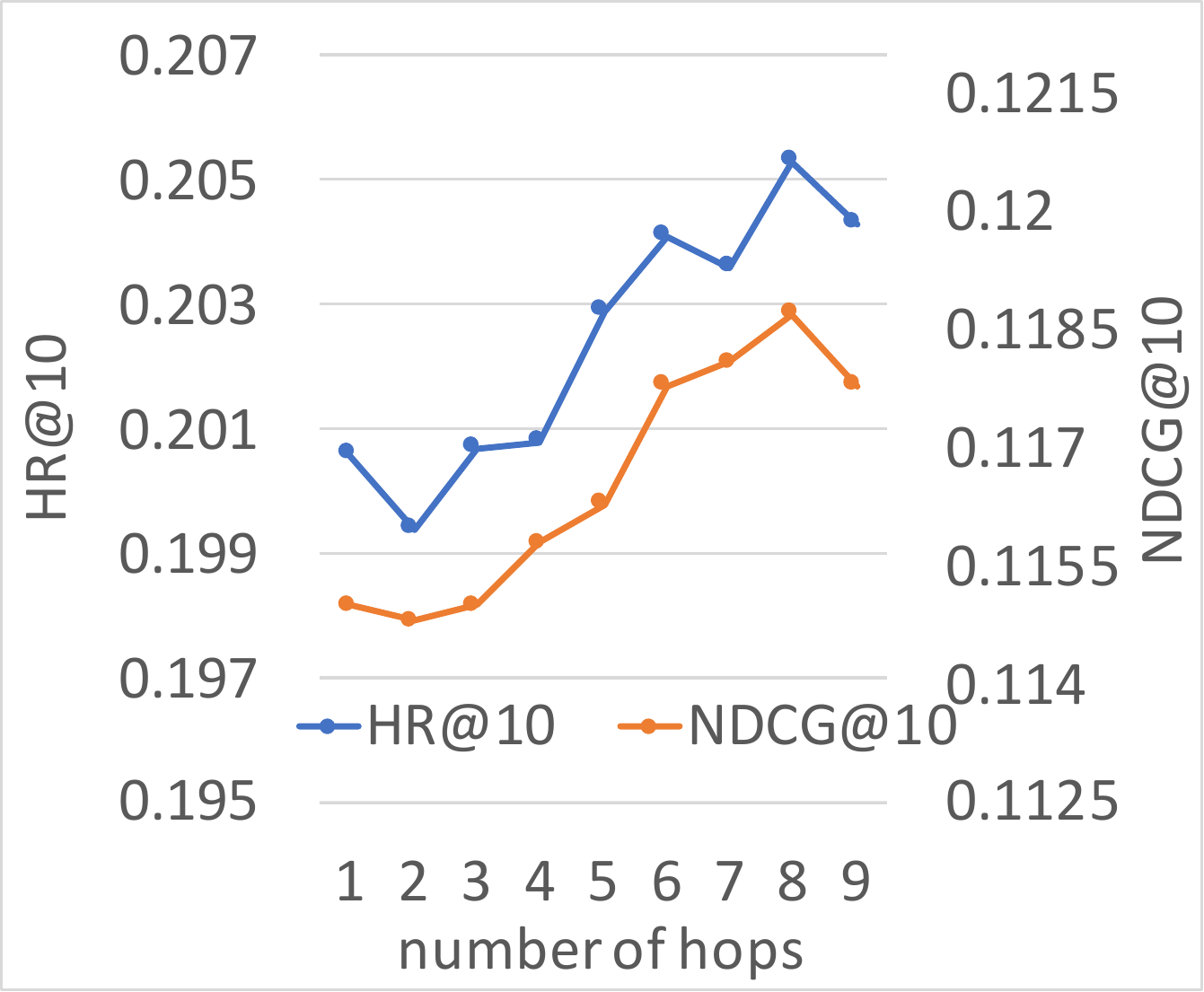}
    }
    \vspace{0pt}
        \subfigure[HR@10 and NDCG@10 on Amazon Electronics]{
        \includegraphics[width=0.1545\textwidth, height=0.13\textwidth]{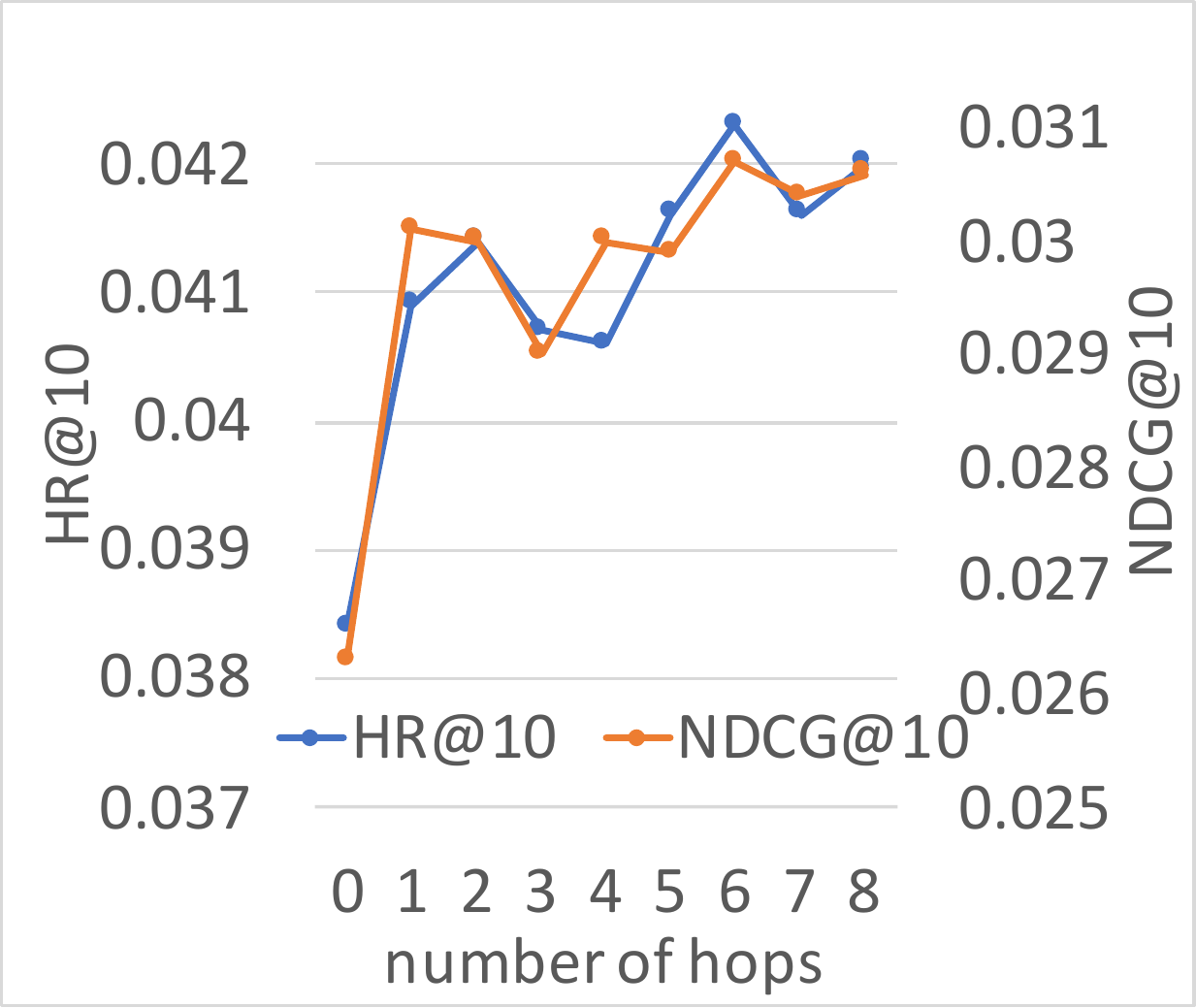}
    }
    \vspace{0pt}
        \subfigure[HR@10 and NDCG@10 on Amazon CDs \& Vinyl]{
        \includegraphics[width=0.1545\textwidth, height=0.13\textwidth]{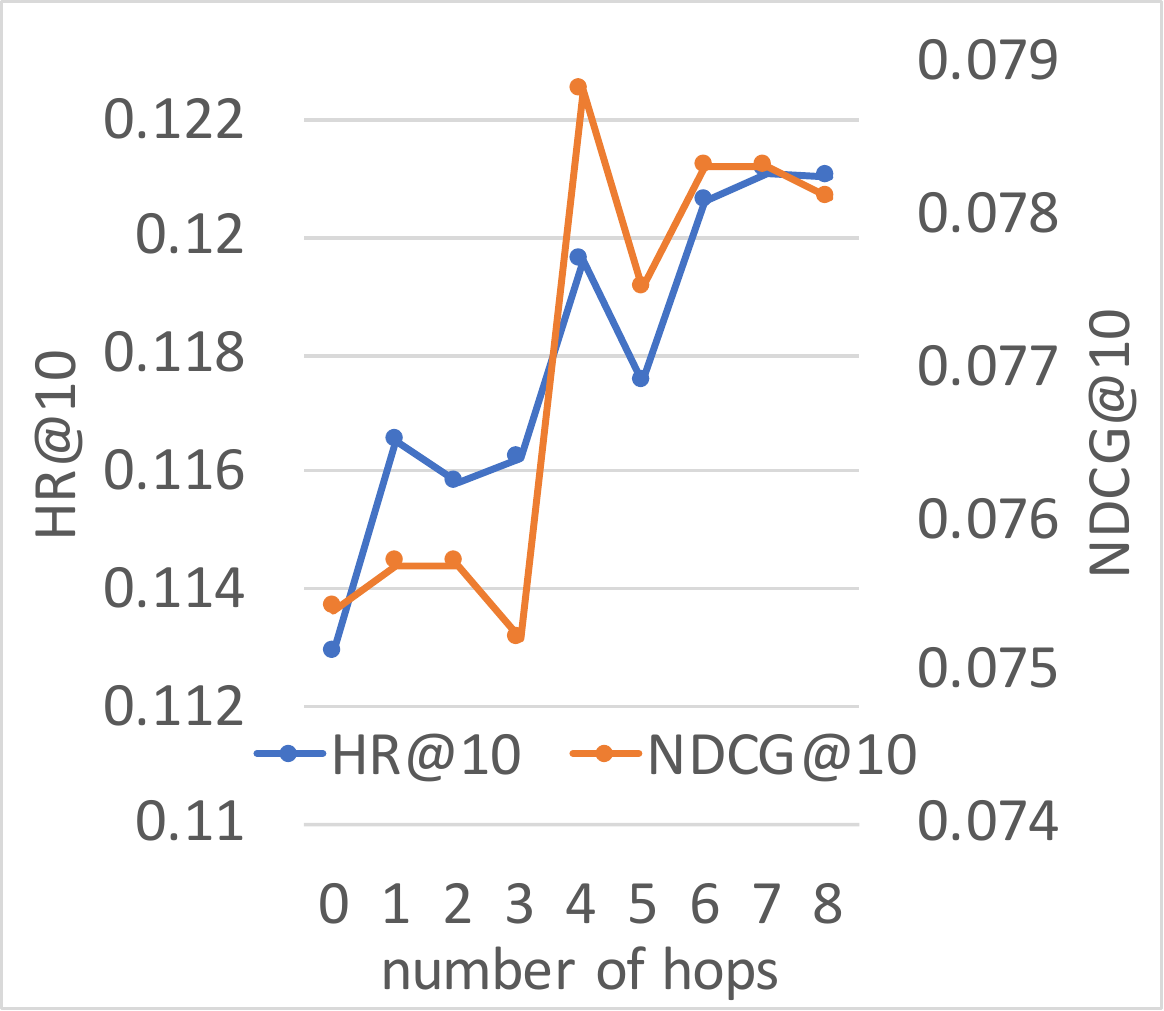}
    }
    \vspace{0pt}
        \subfigure[HR@10 and NDCG@10 on Amazon Movies \& TV]{
        \includegraphics[width=0.1545\textwidth, height=0.13\textwidth]{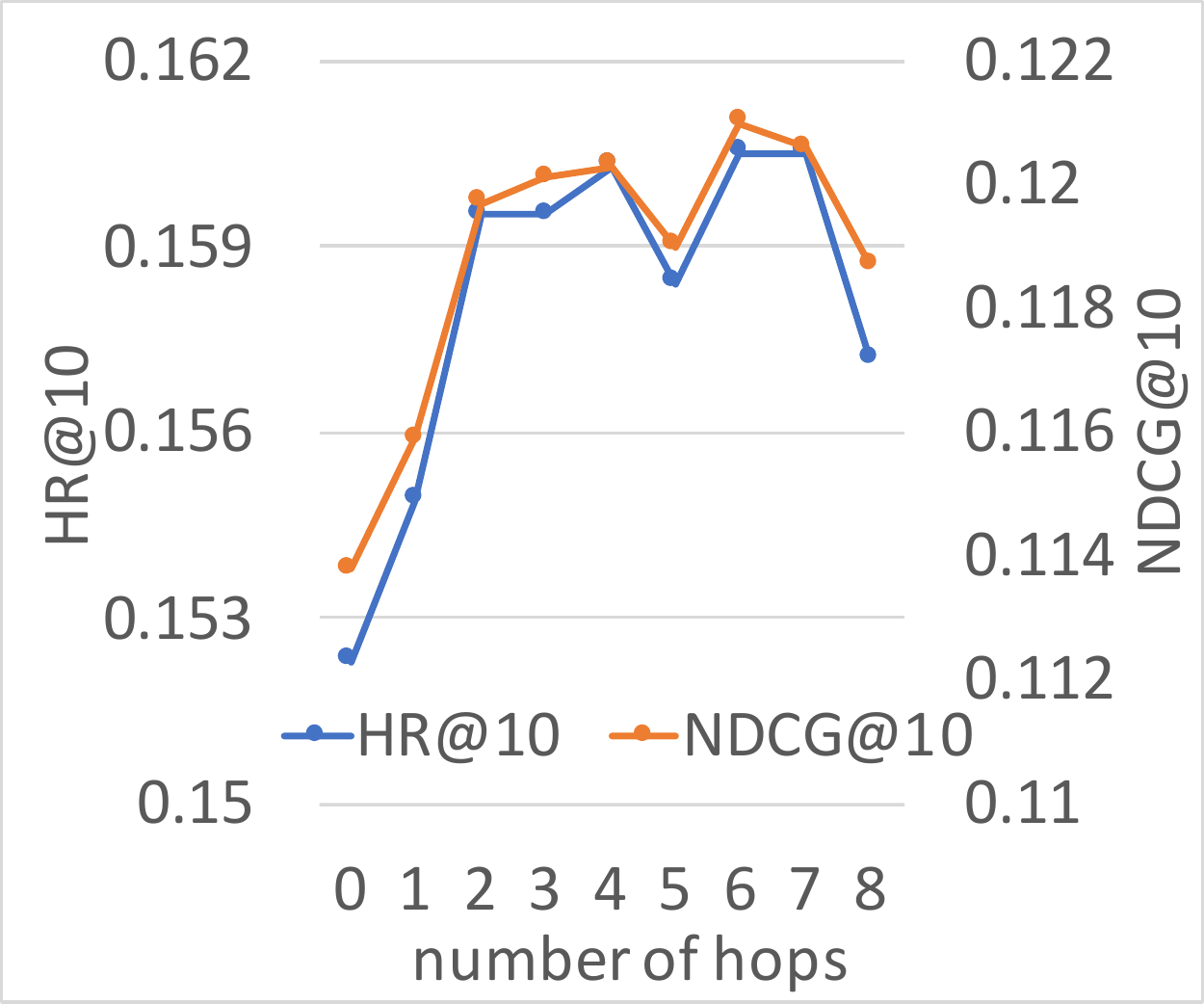}
    }
    \vspace{0pt}
        \subfigure[HR@10 and NDCG@10 on Yoochoose]{
        \includegraphics[width=0.1545\textwidth, height=0.13\textwidth]{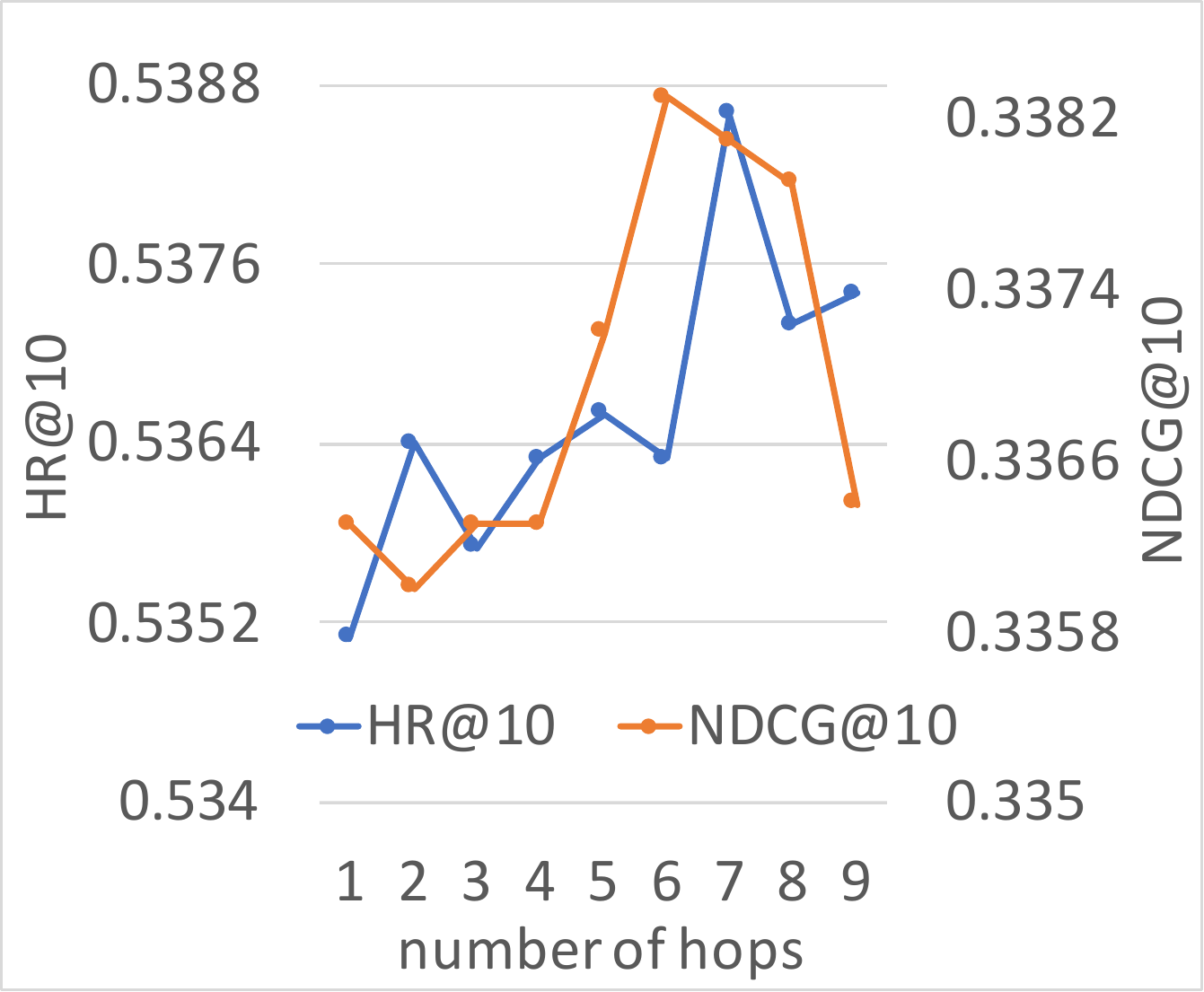}
    }
    \vspace{0pt}
        \subfigure[HR@10 and NDCG@10 on Ali Mobile]{
        \includegraphics[width=0.1545\textwidth, height=0.13\textwidth]{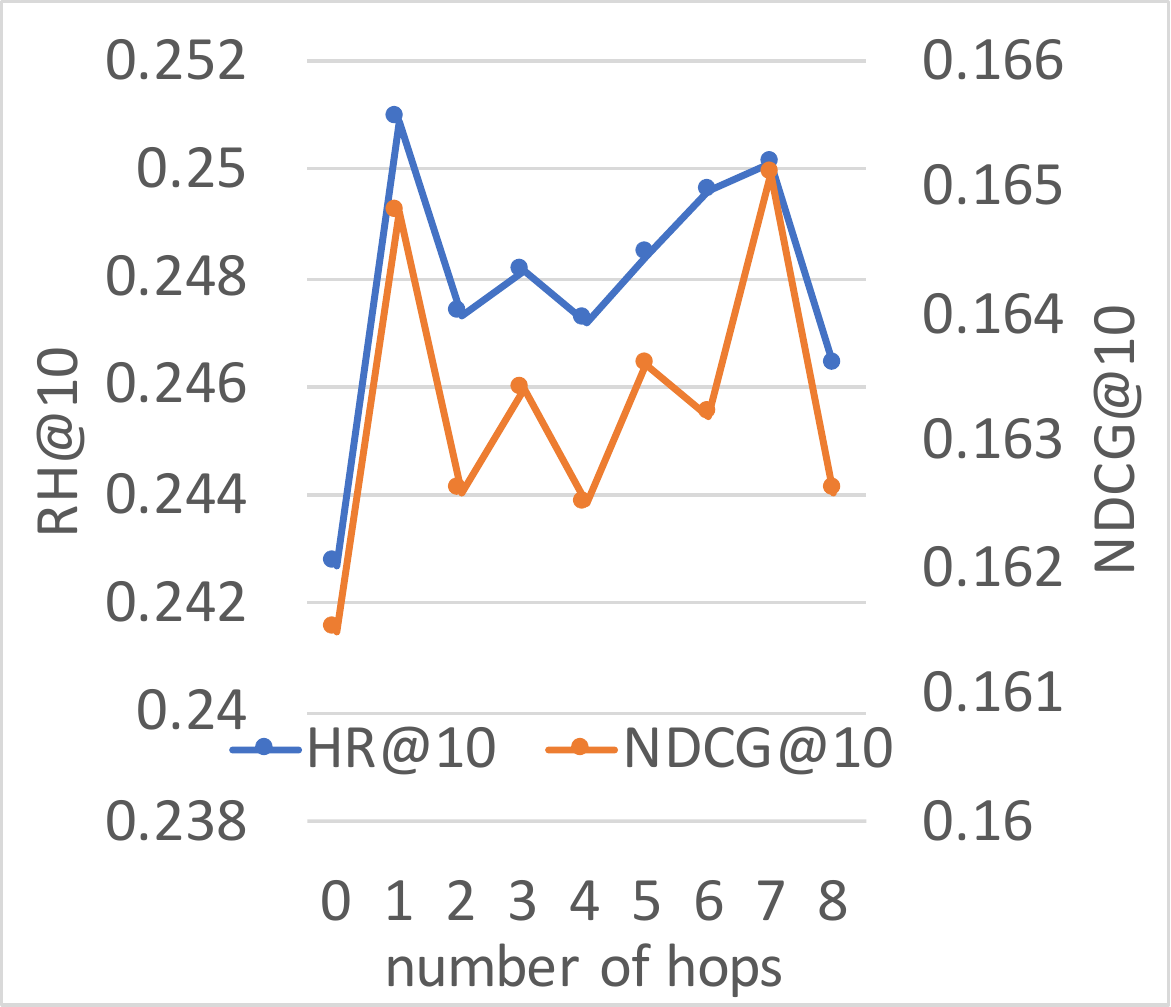}
    }
    \caption{The performance comparison among MTAMs with different number of hops on six datasets, where MTAM with 0 hop is T-GRU.}
    \label{fig:hops} 
\end{figure*} 

(4) We observe that \textbf{MTAM} provides more competitive results than \textbf{NARM++} on five of six datasets, while they are comparable on the left one.
 \textbf{MTAM} and \textbf{NARM++} are both implemented with the proposed time-aware attention and T-GRU.
The differences between them are that \textbf{MTAM} integrate long-term and short-term preferences by a multi-hop memory network, while \textbf{NARM++} uses attention mechanism on the hidden states of RNN to learn the local preferences of users, and combines the local and global preferences with a bi-linear decoder.
The experimental results prove that a multi-hop memory network provides a better way for information fusion.

\vspace{-15pt}
\subsection{Ablation Analysis for \textbf{MTAM}.}
To verify the effectiveness of the proposed time-aware attention, T-GRU and the multi-hop structure of \textbf{MTAM}, we conduct an ablation analysis to demonstrate the contribution of each module. Similar with the experiments of overall performance in Section 6.4, we only report the results of $K = 10$ for all datasets in Table~\ref{tab:Ablation}.
%, and results of different $K$s on Amazon Electronics and Ali Mobile in Figure~\ref{fig:AblationElectronics}, \ref{fig:AblationAli}.

From the results in Table~\ref{tab:Ablation}, 
%and Figure~\ref{fig:AblationElectronics}, \ref{fig:AblationAli}, 
we have some observations:

(1) \textbf{MTAM} defeats \textbf{MTAM with T-SeqRec} for HR@10 on four of the six datasets, while only defeats it for NDCG@10 on four of the six datasets.
Similar results can also be observed in the comparison between \textbf{T-GRU} and \textbf{T-SeqRec} in Table~\ref{tab:performance}.
These results indicate that the proposed \textbf{T-GRU} performs better at recall task than ranking task.
Since the mission of \textbf{MTAM} is Top-K recommendation in the candidate retrieval stage, T-GRU
seems to be the better choice to capture user's short-term intent.

(2) We observe that \textbf{MTAM} performs better than MTAMs based on traditional attention mechanism and GRU unit on five of the six datasets, except that \textbf{MTAM without Time-Aware Attention} dominates other models on Movies \& TV dataset.
These results prove that in most cases the proposed temporal gating methodology improves 
the performance of attentive and recurrent neural models for Top-K recommendation task.

(3) \textbf{MTAM} is obviously more competitive than \textbf{MTAM via T-GRU} and \textbf{MTAM via GRU}. \textbf{MTAM} maintains behavior embeddings in memory, while \textbf{MTAM via T-GRU} and \textbf{MTAM via GRU} store the hidden states of RNN in memory. This result confirms that memory network is more effective to learn long-term dependencies than attentive recurrent networks (e.g. \textbf{NARM}).
An explanation is that RNNs would forcefully summarize the information of all prior behaviors into a hidden state, which make it difficult to assign the credit of each behavior in prediction.

(4) To demonstrate that the proposed time-aware attention is a general improved version of attention mechanism and can be applied in attentive RNNs, self-attention networks and memory networks, we not only equips \textbf{NARM} and \textbf{MTAM} with it, but also update \textbf{SASRec} to a time-aware version.
To our surprise, \textbf{Time-Aware Self Attention} achieves the best performance among all models on two transaction datasets, but loses to \textbf{MTAM} and most of other reassembled MTAMs on all rating datasets. 
This interesting observation first illustrates that the temporal distance is of great importance to attention mechanism when calculating the correlation between two hidden vectors.
Secondly, it implies that time-aware self-attention models may be more competent to handle transaction sequences than attentive recurrent models and memory models.
But on the other hand, attentive recurrent models and memory models are probably better choices to deal with rating sequences.

\subsection{Influence of Multiple Hops} 

Finally, we are curious about whether reading user memory for multiple hops helps to improve the performance of \textbf{MTAM}.
We study the performance of \textbf{MTAM} for HR@10 and NDCG@10 by tuning the number of hops in the range of $0 \sim 8$. 

Results are shown in Figure~\ref{fig:hops}.
We observe that the \textbf{MTAM}s with multiple hops preform better than the \textbf{MTAM} with one hop on five of six datasets.
The only exception is a transaction dataset, Ali Mobile, in which the average length of user behavior sequences is much longer than the other five datasets (shown in Table~\ref{tab:Statistics}).
On Ali Mobile dataset, the single-hop \textbf{MTAM} improves \textbf{T-GRU} obviously, but fails to be further strengthened by performing the reading operation for multiple hops.
This result indicates that a deeper memory network may be not suitable for all datasets.
But, overall, multiple hops improve the performance of \textbf{MTAM}. %, but we see that the effectiveness of multiple hops are quite different in each datasets.
The best number of hops varies from one dataset to another, and in most cases, \textbf{MTAM} needs to read user memory for more than four hops in our experiments.

\section{Conclusions and future works}
In this paper, we proposed a novel Multi-hop Time-aware Attentive Memory (MTAM) network for the task of sequential recommendation.
We first updated attention mechanism and recurrent unit with a new temporal gate to capture the temporal context of user behaviors. 
Then we encoded user's short-term intent with the proposed T-GRU and maintained user's long-term records in memory.
Finally, the user modeling procedure can be viewed as a decision making progress by reading user memory for multiple hops based on the short-term intent.
The experimental results clearly demonstrated the effectiveness of MTAM in Top-K recommendation task and the general improvement of the proposed temporal gating methodology to update the traditional attentive mechanism and recurrent unit for user modeling.

In the experiment, we see a great potential of integrating the temporal gate and self-attentive neural networks.
Compared with NLP tasks, the temporal information are much more important to learn the dependencies between user behaviors than words.
We plan to explore how to take full advantage of temporal information to improve self-attentive neural recommenders in the further.
\vspace{-5pt}
\section{Acknowledgments}
This work was supported by National Key R\&D Program of China (No. 2017YFC0803700), NSFC grants (No.  61532021 and 61972155), Shanghai Knowledge Service Platform Project (No. ZF1213).
\vspace{-5pt}

%\vspace{-10pt}
%%
%% The next two lines define the bibliography style to be used, and
%% the bibliography file.
\bibliographystyle{ACM-Reference-Format}
\bibliography{kdd20}

%%
%% If your work has an appendix, this is the place to put it.

\end{document}